\documentclass{openjournal}
\usepackage{natbib}
\usepackage{academicons}
\usepackage{fontawesome}
\usepackage{xspace}

\usepackage{bm}

\renewcommand{\it}[1]{\textit{#1}}
\usepackage[dvipsnames]{xcolor}
\usepackage{enumitem}
\usepackage{graphicx}	
\usepackage{amsmath}	
\usepackage[linktocpage=true]{hyperref}
\usepackage{subfigure}
\usepackage{setspace}
\usepackage{verbatim}
\usepackage{orcidlink}
\usepackage{makecell}
\usepackage{changepage}
\usepackage{algorithm}
\usepackage{algorithmic}
\usepackage[toc,page]{appendix}
\usepackage{lineno}
\usepackage{beramono}
\usepackage{listings}
\newcommand{\githubmaster}{\href{https://github.com/EdwardBerman/cosmo-corr}{\faGithub}\xspace}

\lstdefinelanguage{Julia}%
  {morekeywords={abstract,break,case,catch,const,continue,do,else,elseif,%
      end,export,false,for,function,immutable,import,importall,if,in,%
      macro,module,otherwise,quote,return,switch,true,try,type,typealias,%
      using,while},%
   sensitive=true,%
   alsoother={$},%
   frame=single,
   morecomment=[l]\#,%
   morecomment=[n]{\#=}{=\#},%
   numbers=left,                        
    numberstyle=\tiny\color{gray},       
   morestring=[s]{"}{"},%
   morestring=[m]{'}{'},%
}[keywords,comments,strings]%

\renewcommand{\arraystretch}{2}

\usepackage[xindy]{glossaries-extra}
\makeglossaries
\GlsXtrEnablePreLocationTag{Page }{Pages }

\definecolor{vlgray}{RGB}{245,245,245}
\definecolor{primary}{RGB}{255,140,0}
\definecolor{bg}{RGB}{250,250,250}
\definecolor{lgreen}{RGB}{184, 209, 126}
\definecolor{dgreen}{RGB}{112, 163, 0}
\definecolor{lorange}{RGB}{255, 199, 128}
\definecolor{dorange}{RGB}{255, 145, 0}
\definecolor{lblue}{RGB}{111, 181, 189}
\definecolor{dblue}{RGB}{0, 95, 117}

\definecolor{darknavyblue}{RGB}{0, 120, 140} 
\definecolor{darkpurple}{RGB}{75, 0, 130} 

\hypersetup{
    colorlinks=true,
    linkcolor=dblue,
    citecolor=NavyBlue,
    urlcolor=darkpurple,
    pdftitle={Soft-Clustering},
    pdfpagemode=FullScreen,
    }

\begin{document}
\shorttitle{On Soft Clustering For Correlation Estimators}
\shortauthors{Berman et al.}

\title{On Soft Clustering for Correlation Estimators}

\author{\href{https://ebrmn.space/}{Edward Berman}$^{1,2, 18~*}$}
\author{Sneh Pandya$^{1,3}$}
\author{Jacqueline McCleary$^{1~*}$}

\author{Marko Shuntov$^{4,5}$}

\author{Caitlin Casey$^{4,6, 17}$}

\author{Nicole Drakos$^{7}$}

\author{Andreas Faisst$^{8}$}

\author{Steven Gillman$^{4,9}$}

\author{Ghassem Gozaliasl$^{10,11}$}

\author{Natalie B. Hogg$^{12}$}

\author{Jeyhan Kartaltepe$^{13}$}

\author{Anton Koekemoer$^{14}$}

\author{Wilfried Mercier$^{15}$}

\author{Diana Scognamiglio$^{16}$}

\author{COSMOS-Web: The JWST Cosmic Origins Survey}

\affiliation{$^{1}$Department of Physics, Northeastern University, Boston, MA 02115, USA}
\affiliation{$^{2}$Department of Mathematics, Northeastern University, Boston, MA 02115, USA}

\affiliation{$^{3}$NSF AI Institute for Artificial Intelligence and Fundamental Interactions (IAIFI)}

\affiliation{$^{4}$Cosmic Dawn Center (DAWN), Denmark} 
\affiliation{$^{5}$Niels Bohr Institute, University of Copenhagen, Jagtvej 128, DK-2200, Copenhagen, Denmark}

\affiliation{$^{6}$Department of Astronomy, The University of Texas at Austin, Austin, TX, USA}

\affiliation{$^{7}$Department of Physics and Astronomy, University of Hawaii, Hilo, 200 W Kawili St, Hilo, HI 96720, USA}

\affiliation{$^{8}$Caltech/IPAC, 1200 E. California Blvd., Pasadena, CA 91125, USA}

\affiliation{$^{9}$DTU-Space, Technical University of Denmark, Elektrovej 327, DK-2800 Kgs. Lyngby, Denmark}

\affiliation{$^{10}$Department of Computer Science, Aalto University, P.O. Box 15400, FI-00076 Espoo, Finland}

\affiliation{$^{11}$Department of Physics, University of Helsinki, P.O. Box 64, FI-00014 Helsinki, Finland}

\affiliation{$^{12}$Laboratoire Univers et Particules de Montpellier, CNRS \& Université de Montpellier, Parvis Alexander Grothendieck, Montpellier, France 34090}

\affiliation{$^{13}$Laboratory for Multiwavelength Astrophysics, School of Physics and Astronomy, Rochester Institute of Technology, 84 Lomb Memorial Drive, Rochester, NY 14623, USA}

\affiliation{$^{14}$Space Telescope Science Institute, 3700 San Martin Drive, Baltimore, MD 21218, USA}

\affiliation{$^{15}$Aix Marseille University, CNRS, CNES, LAM, Marseille, France}

\affiliation{$^{16}$Jet Propulsion Laboratory, California Institute of Technology, 4800, Oak Grove Drive - Pasadena, CA 91109, USA}

\affiliation{$^{17}$Department of Phyiscs, University of California Santa Barbara, CA,93106, CA}

\affiliation{$^{18}$Visiting Scientist, Center for Astrophysics Harvard and Smithsonian, AstroAI}

\thanks{$^\star$ E-mail: \nolinkurl{{berman.ed,  j.mccleary}@northeastern.edu}}

\begin{abstract}
Properly estimating correlations between objects at different spatial scales necessitates $\mathcal{O}(n^2)$ distance calculations. For this reason, most widely adopted packages for estimating correlations use clustering algorithms to approximate local trends. However, methods for quantifying the error introduced by this clustering have been understudied. In response, we present an algorithm for estimating correlations that is probabilistic in the way that it clusters objects, enabling us to quantify the uncertainty caused by clustering simply through model inference. These soft clustering assignments enable correlation estimators that are theoretically differentiable with respect to their input catalogs. Thus, we also build a theoretical framework for differentiable correlation functions and describe their utility in comparison to existing surrogate models. Notably, we find that repeated normalization and distance function calls slow gradient calculations and that sparse Jacobians destabilize precision, pointing towards either approximate or surrogate methods as a necessary solution to exact gradients from correlation functions. To that end, we close with a discussion of surrogate models as proxies for correlation functions. We provide an example that demonstrates the efficacy of surrogate models to enable gradient-based optimization of astrophysical model parameters, successfully minimizing a correlation function output. Our numerical experiments cover science cases across cosmology, from point spread function (PSF) modeling efforts to gravitational simulations to galaxy intrinsic alignment (IA). \githubmaster
\end{abstract}


\maketitle


\begin{figure*}
\centering 
    \subfigure[An overview of our model uncertainty experiment. We repeatedly cluster using a probabilistic algorithm and look at the resulting standard deviation of the resulting correlations.]{
    \includegraphics[width=1\textwidth]{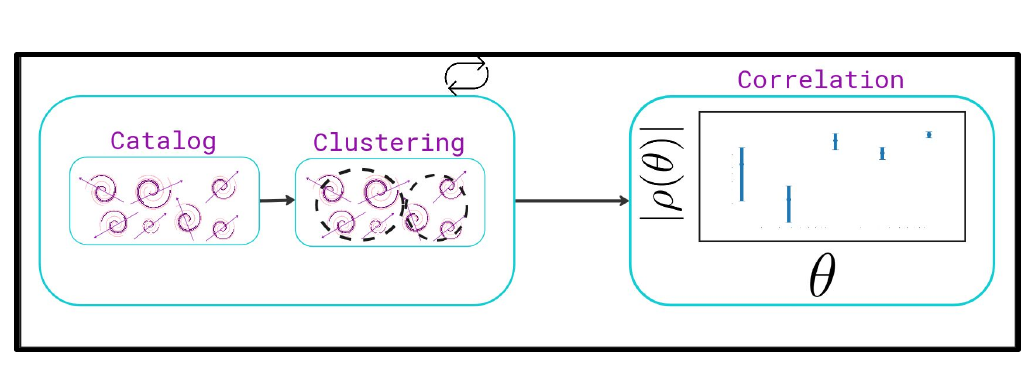}
    
    } 
    \subfigure[An overview of our differentiability experiment. We forward model a gravitational simulation and show that we can differentiate through the estimator even with the clustering approximation.]{
    \includegraphics[width=1\textwidth]{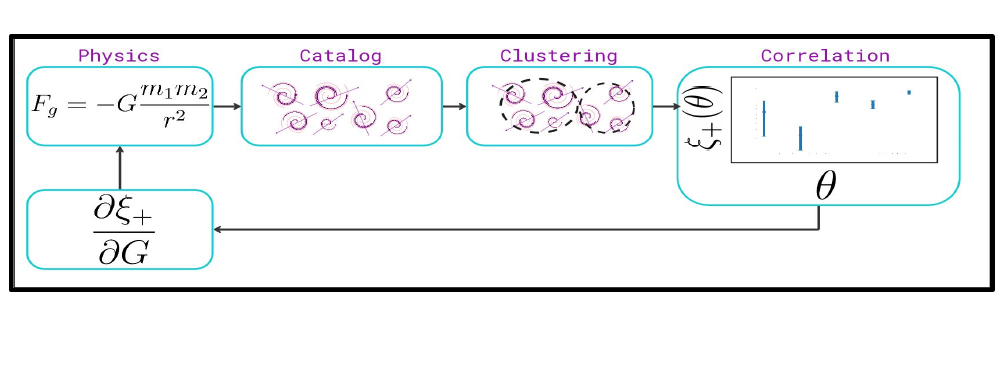}}
         \subfigure[An overview of our surrogates experiment. We exploit the differentiability of neural networks to enable Hamiltonian Monte Carlo for fast posterior sampling of the IA parameters most likely to minimize a correlation function.]{
    \includegraphics[width=1\textwidth]{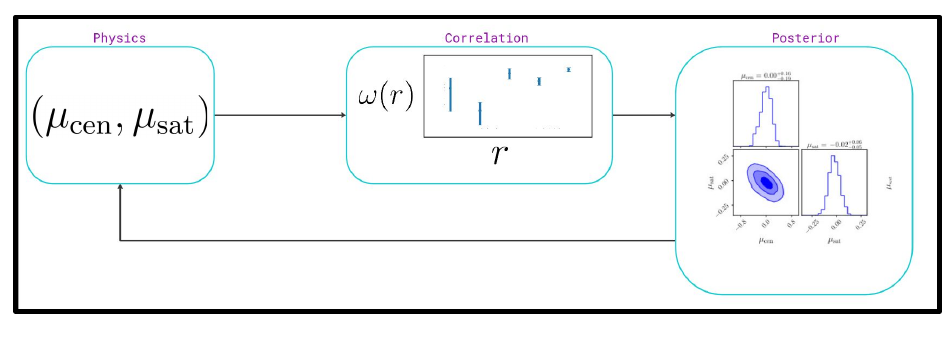}}
    
     \vspace{1em}
     \caption{Outline of the three experiments conducted in this paper. The top outlines our model uncertainty experiment (\S\ref{sec:modeluq}), the middle outlines our differentiability experiment (\S\ref{sec:differentiability}), and the bottom outlines our surrogate experiment (\S\ref{sec:surrogates}).}
    \label{fig:process}
\end{figure*}

\newpage
\section{Introduction} \label{sec:intro}
Two- and three-point correlation functions ($2/3$PCF) are widely used in cosmology. Algorithms to compute these correlation functions first cluster their objects into representative points to ease the computational load of computing $\mathcal{O}(n^2)$ pairwise distances. To numerically compute the correlation function, the pairwise distance between all of the points is then computed, and an estimator is computed on all of the points in each distance bin. 

A long-standing goal in the computational sciences is to reconcile epistemic (model) uncertainties from aleatoric (data) uncertainties \citep{hullermeier2021aleatoric, osband2023epistemic}. Aleatoric uncertainties describe uncertainty that is inherent to a statistic. This can be due to data quality limitations or inherent stochasticity of the quantity one is trying to model. On the other hand, epistemic uncertainties reflect the inability to generalize from a finite data set due to inherent limitations of the model. In the case of $2$PCFs and $3$PCFs, the epistemic uncertainty is often overlooked: the best assignment of an object to a given cluster is often unclear, as data points that are ``in-between'' two or more learned cluster centers can reasonably be chosen to fall in any of the clusters. This ambiguity introduces an epistemic uncertainty, as the model will struggle to appropriately assign objects to representative clusters without enough data. With more data points, it is easier to identify the most appropriate way to cluster the objects and thus epistemic uncertainty should decrease in the main, though individual data points will still face ambiguity. Prior works usually assume this clustering uncertainty to be trivial, arguing that for a significantly large sample, the number of overestimated and underestimated distances are roughly equal, and so the errors tend to cancel out \citep{jarvis2004skewness}. For example, this was assumed for the Dark Energy Survey (DES) point spread function (PSF) modeling efforts \citep{jarvis2021dark}. A PSF is an impulse response of an optical system to light, and the quality of a PSF model is often assessed with $\rho$ statistics \citep{rowe2010improving,vogelsberger2016ethos}, which describe correlations in size and shape residuals of a PSF model relative to observed stars. These empirical PSF models are often trained using point source catalogs that are generated by SourceExtractor++ \citep{bertin1996sextractor} and filtered to be within a given size/magnitude range. DES covered a vast survey area and included millions of objects which were used to be fit and evaluate PSF models \citep{DES_Y3_PIFF, jarvis2021dark}. With so much data at hand, it was fair to assume that the number of overestimated and underestimated distances were roughly equal, and that downstream modeling errors from clustering were negligible. Surveys such as DES therefore look exclusively at uncertainties quantified via bootstrapping or jackknife \citep{jarvis2021dark, TreeCorr}, which captures only the model's robustness to individual data points. This is closer in nature to being an aleatoric uncertainty, and we will refer to it as aleatoric uncertainty throughout. 

\begin{figure}
    \centering
    \includegraphics[width=0.7\linewidth]{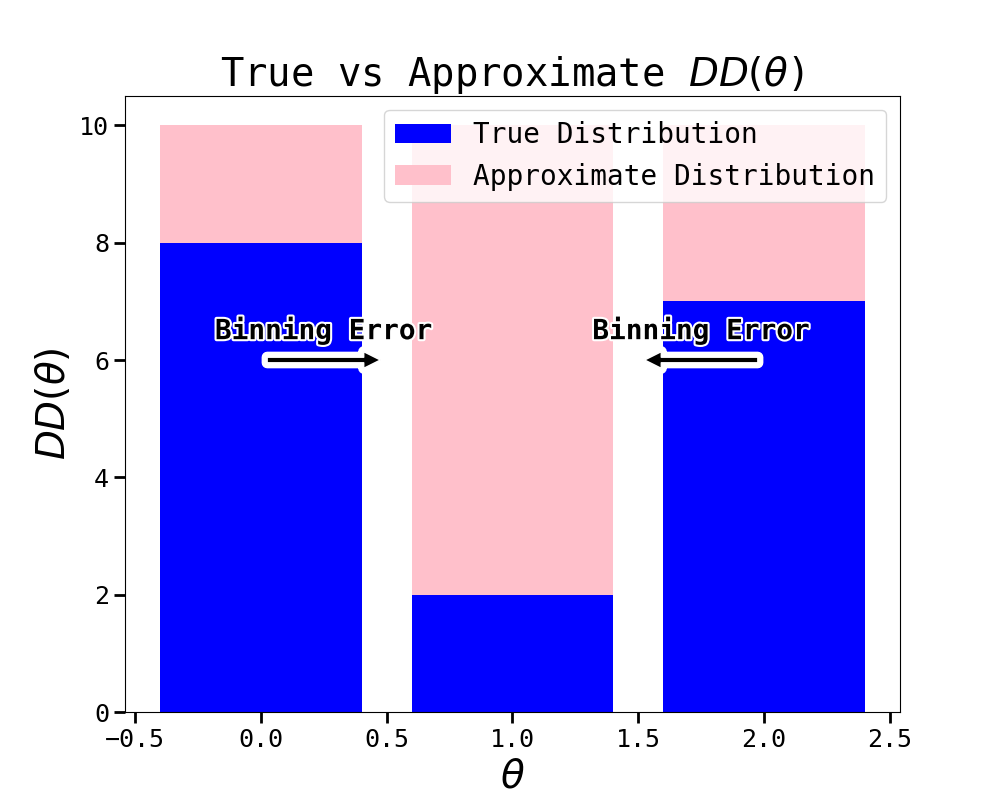}
    \includegraphics[width=0.7\linewidth]{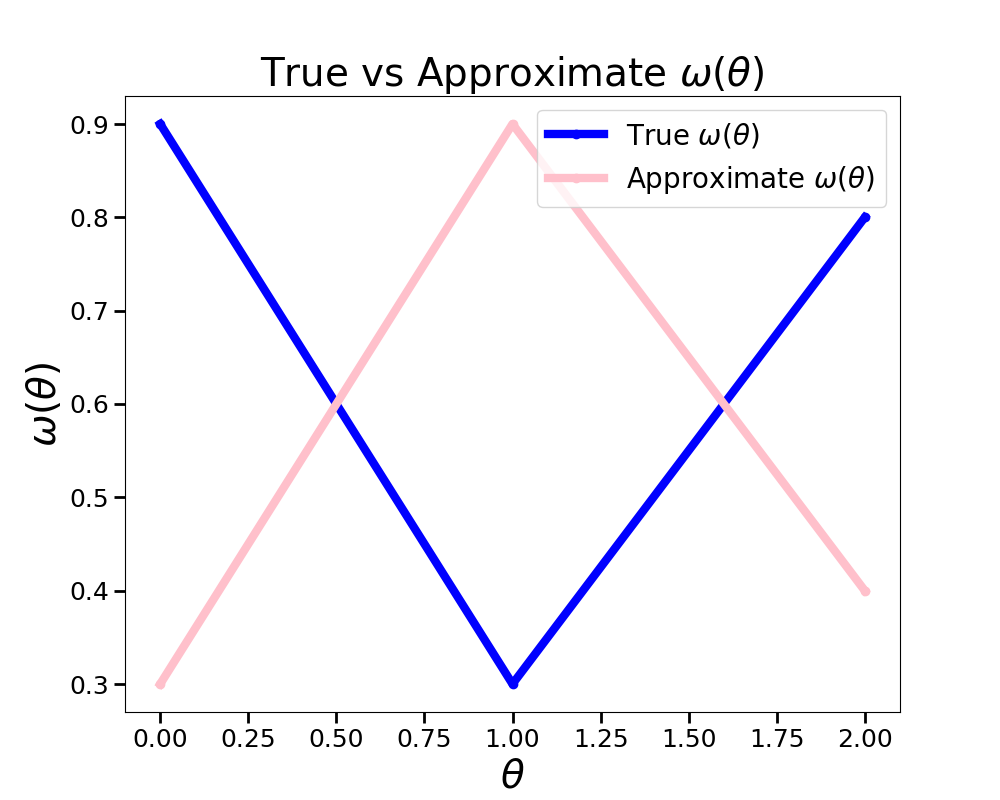}
    \caption{A toy example of a clustering approximation causing a downstream error in the estimation of the $\omega(\theta)$ position-position auto-correlation function.}
    \label{fig:toy-ex}
\end{figure}

This work presents a method to quantify epistemic uncertainty in regimes where one cannot make the same assumptions as \cite{jarvis2004skewness} due to the small number of data points available. For instance, clustering errors cannot be assumed to be negligible for the NIRCam PSF modeling efforts with COSMOS-Web \citep{casey2023cosmos}, where the number of point sources is on the order of a few $100$ across the entire field of view \citep{berman2024efficientpsfmodelingshoptjl}. Even through the TreeCorr algorithm \citep{jarvis2004skewness} makes sure distance offsets are less than a bin width, results can still be heavily biased. A toy example of this with the position-position autocorrelation is presented in Figure \ref{fig:toy-ex}. The unbiased estimator for this correlation is the Landy--Szalay estimator \citep{landy1993bias} shown in Equation \ref{eq:landy}.

\begin{equation}
    \omega (\theta) = \frac{DD(\theta) - 2DR(\theta) + RR(\theta)}{RR(\theta)}
    \label{eq:landy}
\end{equation}
Equation \ref{eq:landy} assumes access to one real observation of galaxies as well as an additional catalog of galaxies randomly distributed on the sky. The terms in Equation \ref{eq:landy} are $DD(\theta)$,  the number of real galaxy pairs separated by an angular distance $\theta$, $RR(\theta)$, the number of random galaxy pairs separated by a distance $\theta$, and $DR(\theta)$, the number of pairs between one real galaxy and one random galaxy separated by a distance $\theta$. Since the estimator is directly proportional to the number of galaxies that are measured to be $\theta$ arcmins apart through $DD(\theta)$, even a single bin offset can heavily skew a measurement. While $\rho$ statistics themselves are not directly related to the number of objects in a single distance bin, the example indicates that binning offsets can constructively bias the result in a given bin.

The core method of this work is to recast object-cluster assignment probabilistically. In this way, we study how the uncertainty related to the clustering of objects can propagate through to the output. Our findings suggest that there are equal amounts of data and model uncertainty in the case of computing $\rho$ statistics. This further implies that one should skip the clustering step to mitigate the clustering uncertainty. 

We also observe that the soft or probabilistic assignment of galaxies to cluster centers makes our estimator theoretically differentiable with respect to its input catalog, and by extension the parameters of any differentiable forward model that produces an input catalog. Differentiability in this context is highly desirable, as it allows one to relate correlations to astrophysical model parameters through gradient-based optimization \citep[for related work, see][]{lanzieri2023forecasting, cuesta2024point, lanusse2023dawes,campagne2023jax, jagvaral2024geometric, 2025AAS, cosmo-inform, halverson2024generalitypersistencecosmologicalstasis}. \cite{halverson2024generalitypersistencecosmologicalstasis} in particular describes a pipeline analogous to what we would like to achieve. They optimize cosmological model parameters to maximize the number of extended cosmological epochs where the relative abundances of different $\Omega_i$ remain unchanged, referred to as periods of cosmological stasis \citep{dienes2022stasis}. In our case, we are interested in what kind of implicit cosmological parameters maximize or minimize different types of two and three-point correlations (cf. Figure \ref{fig:process}). Motivated by this desire to maximize or minimize correlation functions, we provide a discussion of the gradient related properties of our algorithm. In contrast to our method, clustering schemes that use hard assignments (i.e. every object has one assignment with unit probability) often depend on functions like \textsc{Median} or \textsc{argmin}, which results in derivatives that are either identically zero or ill-defined. We further explore the differentiability of our clustering scheme with an experiment. In our experiment, we use a gravitational simulation as a forward model with the goal of differentiating a correlation function with respect to the gravitational constant, $G$. Our findings motivate the use of surrogate models: learned functions that approximate the relationship between astrophysical model parameters and observable data. Recent studies demonstrate their ability to directly model two-point statistics from astrophysical model parameters \citep{pandya2024learning, pandya2025iaemulearninggalaxyintrinsic}, making them ideal for our purpose. Accordingly, we extend the work of \cite{pandya2024learning, pandya2025iaemulearninggalaxyintrinsic}, using the Intrinsic Alignment Emulator (IAEmu) surrogate model to optimize seven halo-based modeling parameters to minimize a correlation function. In doing so, we illustrate the effectiveness of surrogate models to  enable gradient-based optimization of astrophysical model parameters, which in turn allows us to study what physical circumstances cause correlations to be maximized or minimized.

Our paper executes three experiments that are designed to explore three themes: model uncertainty, differentiability through an estimator, and differentiability though a surrogate. 

\begin{enumerate}
    \item \textbf{Model Uncertainty and PSFs ($\S$\ref{sec:modeluq}):} For this experiment, we use ShOpt.jl \citep{berman2024efficientpsfmodelingshoptjl, Berman2024} to produce PSF models for the COSMOS-Web NIRCam images \citep{casey2023cosmos}. Using our proposed algorithm, we then compute $\rho$ statistics. Finally, we assess the epistemic uncertainty associated with our model and compare it to the uncertainty we would get with traditional bootstrapping techniques. Our PSF modeling experiment is geared towards observers, who do not have access to differentiable forward models of PSF residuals. 
    \item \textbf{Differentiability and Gravity Simulations ($\S$\ref{sec:differentiability}):} For this experiment, we forward model gravitational interactions using a series of ordinary differential equations. We then compute correlation functions on the resulting catalog and test different techniques for differentiating the correlation function outputs with respect to the cosmological parameters that determined the simulation. Our gravitational simulation experiment is designed with theorists in mind. Theorists can produce multiple realizations of their model through simulations that are derived from different cosmological parameters. Here, we assess the feasibility of differentiating a correlation function with respect to $G$.  
    \item \textbf{Surrogates and Galaxy Intrinsic Alignment ($\S$\ref{sec:surrogates}):} For this experiment, we use an existing surrogate model to relate astrophysical model parameters to IA \citep{pandya2024learning, pandya2025iaemulearninggalaxyintrinsic}. Using this surrogate, we use Hamiltonian Monte Carlo (HMC) to find posterior distributions over astrophysical model parameters that are likely to correspond to correlations of zero. Our IA experiment shows how to compensate for forward models and estimators that are not easily differentiated, and are again geared towards theorists who are able to realize many instances of their forward model through inference in order to train a surrogate. Our usage of HMC leverages the differentiability of neural networks to accelerate the sampling procedure in comparison to traditional algorithms like Metropolis--Hastings \citep{betancourt2017conceptual, robert2004metropolis}.
\end{enumerate}

The remainder of the paper is organized as follows: $\S$\ref{sec:notation} discusses the relevant notation and theory for $\rho$ statistics and IA, $\S$\ref{sec:data} follows with an outline of the data products used in our study, $\S$\ref{sec:algo} lays out additional notation as well as details for the algorithm used in this paper, and $\S$\ref{sec:modeluq} -\ref{sec:surrogates} presents the results of the experiments described above followed by our conclusions in $\S$\ref{sec:conclusions}.

\section{Notation and Theory} \label{sec:notation}

In this section we discuss the relevant theory and notation for correlations of PSF size and shape residuals and of IA. We also expand on the theory we use to compute $\sigma_{\text{epistemic}}$ and $\sigma_{\text{aleatoric}}$.

\subsection{PSF Modeling}

A PSF describes the impulse response of an optical system to light. Effects like diffraction, optical aberrations, atmospheric turbulence (if applicable), and telescope jitter are summarized in the telescope’s PSF. For the James Webb Space Telescope (JWST), the PSF provides an obstacle for science goals including the highest resolution dark matter mass maps \citep[][Scognamiglio et al., in prep]{Scognamiglio2024Exploring} and the characterization of over $100$ strong lens systems (Nightingale et al., in prep). $\rho$ statistics are a suite of 2PCFs introduced in \cite{rowe2010improving} and expanded upon in \cite{vogelsberger2016ethos} for characterizing biases in size and shape measurement that arise from PSF modeling.

We adopt the same notation for describing $\rho$ statistics as \cite{jarvis2021dark} and \cite{McCleary_2023}:

\begin{align}
        \rho_1(\theta) &\equiv 
            \left\langle 
            \delta e^*_{\rm PSF}(\bm{x}) 
            \delta e_{\rm PSF}(\bm{x}+\bm{\theta})
            \right\rangle \label{eq:rho1}\\ 
        \rho_2(\theta) &\equiv 
            \left\langle 
            e^*_{\rm PSF}(\bm{x}) 
            \delta e_{\rm PSF}(\bm{x}+\bm{\theta})
            \right\rangle\\
        \rho_3(\theta) &\equiv 
            \left\langle 
            \left( e^*_{\rm PSF}\frac{\delta T_{\rm PSF}}{T_{\rm PSF}}\right)(\bm{x})
            \left(e_{\rm PSF}\frac{\delta T_{\rm PSF}}{T_{\rm PSF}}\right)(\bm{x}+\bm{\theta})
            \right\rangle\\
        \rho_4(\theta) &\equiv 
            \left\langle 
            \delta e^*_{\rm PSF}(\bm{x})
            \left(e_{\rm PSF}\frac{\delta T_{\rm PSF}}{T_{\rm PSF}}\right)(\bm{x}+\bm{\theta})
            \right\rangle\\  
        \rho_5(\theta) &\equiv 
            \left\langle 
            e^*_{\rm PSF}(\bm{x})
            \left(e_{\rm PSF}\frac{\delta T_{\rm PSF}}{T_{\rm PSF}}\right)(\bm{x}+\bm{\theta})
            \right\rangle \label{eq:rho5}
\end{align}
\noindent where $e_{\rm PSF}$ is the ellipticity of the real PSF, i.e., the star ellipticity, $T_{\rm PSF}$ is the size of the real PSF, $\delta e_{\rm PSF}$ is the difference between the ellipticity of the real and model PSFs at position $\bm{x}$, and $\delta T_{\rm PSF}$ is the difference between the sizes of the real and model PSFs at position $\bm{x}$. Brackets denote averages over all pairs within a separation $\bm{\theta}$, and asterisks denote complex conjugates. $T_{\rm PSF}$ is nominally defined as $2\sigma^2$, where $\sigma^2$ is the variance of the best fit elliptical Gaussian to the PSF.

$\rho$ statistics specify the $\xi_+$ weak lensing shear correlations, which are given by

\begin{equation}
    \xi_+(\theta) = \langle \epsilon_{t}(x+\theta) \epsilon_{t}(x) \rangle + \langle \epsilon_{\times}(x+\theta) \epsilon_{\times}(x) \rangle 
\end{equation}

and estimated by 

\begin{equation}
    \hat{\xi}_+(\theta) = \frac{\sum_{i,j} w_i w_j(\epsilon_{t,i} \epsilon_{t,j} + \epsilon_{\times,i} \epsilon_{\times,j})}{\sum_{i,j}w_iw_j}.  \label{eq:estimator}
\end{equation}

where $w_{i/j}$ represents the weight uncertainties of object $(i/j)$, $\epsilon_{t, i/j}$ represents tangential shear of object $(i/j)$ and $\epsilon_{\times, i/j}$ represents the cross shear of object $(i/j)$ \citep{kilbinger2015cosmology,schneider2002analysis}. The complex components of the products in Equations \ref{eq:rho1} - \ref{eq:rho5} are known to average out to zero due to parity symmetry \citep{rowe2010improving, schneider2006gravitational}, and are thus ignored in the estimator. 

\subsection{Galaxy Intrinsic Alignment}

Intrinsic alignment describe correlations between galaxy orientations and the underlying distribution of dark matter where they are embedded; see \cite{lamman2023ia} for a review. The three relevant IA correlation functions are the galaxy position-position $\xi(r)$, position-orientation $\omega(r)$, and orientation-orientation $\eta(r)$ correlation functions, and are estimated according to Equations \ref{eq:xir}-\ref{eq:etar}.

\begin{eqnarray}
    \xi(r) &=& \left\langle \frac{n(r) - \bar{n}(r)}{\bar{n}(r)} \right\rangle \label{eq:xir}\\
    \omega(r) &=& \langle |\hat{e}({\bf x}) \cdot \hat{r}|^2 \rangle -\frac{1}{3}\\
    \eta(r) &=& \langle |\hat{e}({\bf x}) \cdot \hat{e}({\bf x}+{\bf r})|^2 \rangle -\frac{1}{3} \label{eq:etar}
\end{eqnarray}

where $n(r)$ is the number of galaxies separated by distance $r$, $\bar{n}(r)$ is the expected number of galaxies separated by distance $r$ for a random distribution, $x$ is the position vector of a galaxy, and $\hat{e}(x)$ is a 3D orientation unit vector of a galaxy. 

\subsection{Uncertainty Quantification}

Given a probabilistic clustering algorithm for estimating correlations, $\sigma_{\text{epistemic}}$ is computed by estimating the correlation value in each distance bin $n$ times and then computing the standard deviation of the distribution. For computing $\sigma_{\text{aleatoric}}$ via bootstrapping or jackknife, one must first make the probabilistic clustering algorithm deterministic. In this work, we use greedy sampling to accomplish this; that is, each object is always assigned to the cluster with its highest membership probability. In this way, the model uncertainty is isolated from the data uncertainty. To perform bootstrapping, random subsets of the input data is removed over $n$ iterations and the correlation value is estimated in each distance bin. $\sigma_{\text{aleatoric}}$ is then computed via the standard deviation of the output in each distance bin. We choose $n = 10$ samples to obtain the error bars.

Our definitions of aleatoric and epistemic are consistent with what is used in the machine learning literature \citep{hullermeier2021aleatoric}. It is worth noting that in the cosmology literature, these uncertainties are sometimes described as statistical and systematics, e.g. in \cite{freedman2024status}. Following \cite{freedman2024status}, we take our uncertainties to be additive. That is, correlation functions will have some value $\mu \pm \sigma_{\text{aleatoric}} \pm \sigma_{\text{epistemic}}$ in each distance bin. $\mu$ can be obtained from the average of the either the probabilistic or deterministic algorithm samples.

\section{Data} \label{sec:data}

\subsection{Point Source Catalog}

\begin{figure}[!htb]
    \centering
    \includegraphics[width=1\linewidth]{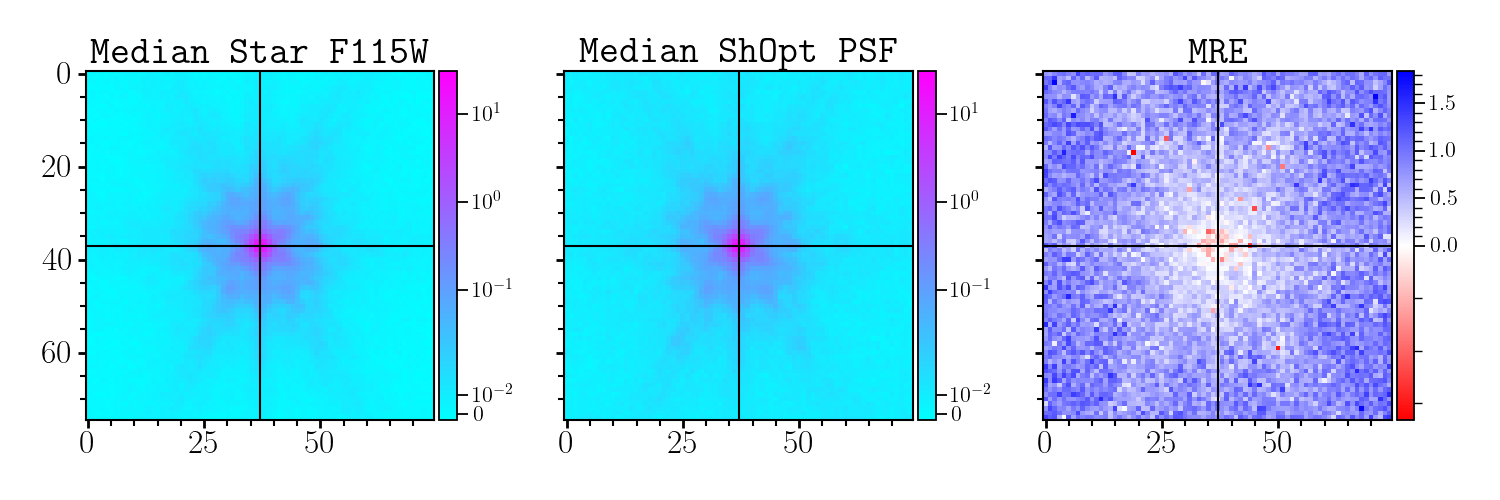}
    \caption{ShOpt PSF fit used in this work. The left panel shows the median of the vignettes. The center panel show the median PSF cutout. The right panel show the average relative error between the vignette cutouts and the PSF cutouts. More on these plots can be found in \cite{berman2024efficientpsfmodelingshoptjl}. }
    \label{fig:PSF}
\end{figure}

\begin{figure}[!htb]
    \centering
    \includegraphics[width=0.5\linewidth]{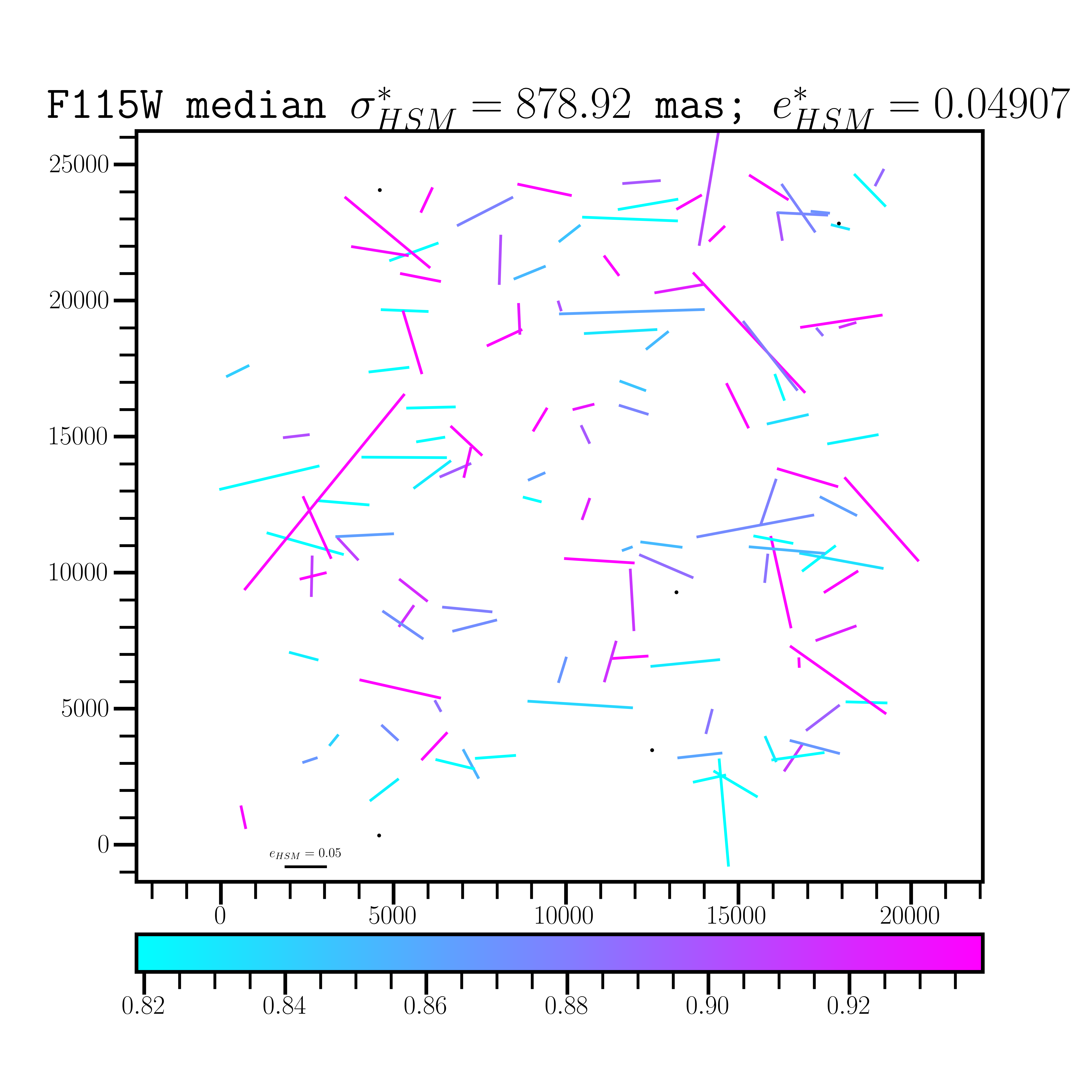}
    \caption{A quiverplot indicating the sizes and shape of the point sources computed via HSM adaptive moments in a single tile of the COSMOS-Web survey. The color bar at the bottom denotes the size of the source, the length of the quiver describes the magnitude of ellipticity, and the orientation and position describes its measured orientation and position on the image plane. The median size and ellipticity of point sources in terms of HSM moments is given at the top of the figure.}
    \label{fig:quiver}
\end{figure}

The point sources used for generating our PSF models come from the SE$++$ \citep{bertin1996sextractor} COSMOS-Web catalog \citep[][Shuntov et al. in prep]{casey2023cosmos}. We restrict ourselves to imaging from the F$115$W NIRCam \citep{2003SPIE,2005SPIE,2012SPIE} wavelength filter, as the PSFs in this wavelength are the closest to being well-approximated by a Gaussian \citep{berman2024efficientpsfmodelingshoptjl}. $\rho$ statistics and the related Hirata--Seljak--Mandelbaum (HSM) adaptive moments \citep{hirata2003shear,mandelbaum2005systematic} are based off of the Gaussian approximation of the PSF, and so the use of $\rho$ statistics in our experiments is most appropriate in this wavelength. PSF models are obtained using ShOpt.jl \citep{berman2024efficientpsfmodelingshoptjl, Berman2024}. Our choice of PSF fitter is discussed in more detail in Appendix \ref{sec:psf_appendix}. Our ShOpt fits can be seen in Figure \ref{fig:PSF}. The ellipticites of the objects in our data set are also visualized in Figure \ref{fig:quiver}. We use the same configuration file as found in the appendix of \cite{berman2024efficientpsfmodelingshoptjl}. Our data set statistics for PSF modeling are summarized in Table \ref{tab:data}.

\begin{table}[!htb]
    \centering
    \begin{tabular}{cccc}
       Training Stars  & Validation Stars & Filter & Survey Area (deg$^2$) \\ \hline
        6333 & $610$ & F$115$W & $0.59$\\
    \end{tabular}
    \caption{Data set statistics for PSF catalog}
    \label{tab:data}
\end{table}

\subsection{Simulated Galaxy Catalog}

There are plenty of widely-used physically realistic N-body simulations, using publically available codes such as Gadget-2/3 \citep{springel2005cosmological}, ART \citep{kravtsov1997adaptive}, RAMSES \citep{teyssier2002cosmological}, and Enzo \citep{brummel2019enzo, bryan2014enzo}, see \cite{vogelsberger2020cosmological} for a review. However, this work aims only to have a simple differentiable forward model. Instead of using Gadget-2 or similar tools, we write our simulations from scratch in Julia and Python and time evolve objects according to ordinary differential equations from Newtonian mechanics. Our main implementation uses Jax \citep{jax2018github} with Diffrax \citep{kidger2021on}. We specify starting positions and velocities for objects $[\vec{x}, \vec{v}]$, which are then evolved via $\frac{d\vec{x}}{dt} = \vec{v}$, $\frac{d\vec{v}}{dt} = \vec{a}$, and $\vec{a} = - \frac{GM}{|\vec{x}|^3}\vec{x}$. We begin by randomly initializing galaxies with some mass $M$, velocity $\vec{v}$, and initial position $\vec{r}$. These galaxies exist in a gravitational field where the galaxy-galaxy gravitational attraction is assumed to be negligible compared to the attraction toward the center of the field. Thus, we can solve each ODE system in Diffrax independently and use the vmap function to efficiently time evolve each object. While our toy simulation is not very physically meaningful, this approach outputs catalogs that are differentiable with respect to the input cosmology, which in this case is just the gravitational constant $G$. The number of galaxies in the simulation can vary, allowing us to study the efficacy of our algorithm as a function of catalog sizes.

\subsection{Intrinsic Alignment Catalog}

This work uses an extensive galaxy catalog that was generated according to seven halo-based modeling parameters using the procedure of \cite{pandya2024learning, pandya2025iaemulearninggalaxyintrinsic}. The halo occupation distribution (HOD) model is used to populate existing catalogs of dark matter-only halos with galaxies \citep{Hearin_2017}. The model is parameterized by five occupation components, $\log M_{\rm min}$, $\sigma_{\log M}$, $\log M_0$, $\log M_1$, and $\alpha$, described in detail in \citep{Zheng_2007}. Briefly, the occupation components parameterize the mean occupation functions given by 
\begin{equation}
    \langle N_{\rm cen}(M) \rangle = \frac{1}{2}\left[1 + \text{erf}\left(\frac{\log M - \log M_{\rm min}}{\sigma_{\log M} }\right) \right]
\end{equation}
and 
\begin{equation}
    \langle N_{\rm sat}(M)\rangle = \frac{1}{2}\left[1 + \text{erf}\left(\frac{\log M - \log M_{\rm min}}{\sigma_{\log M} }\right) \right]\left(\frac{M - M_0}{M_1'}\right)^\alpha
\end{equation}

where \begin{eqnarray}
    \text{erf}(x) = \frac{2}{\sqrt{\pi}}\int_0^x e^{-t^2}dt.
\end{eqnarray}

The contributions of \cite{van2024empirical} introduce a model of IA within a HOD-based framework. Specifically, the IA implementation includes a two-parameter family governing central and satellite alignment strengths ($\mu_{\rm cen}$ and $\mu_{\rm sat}$). The work explores different forms of alignment strength (e.g., constant, radially dependent), and demonstrates that this parameterization can effectively model IA in both dark-matter only simulations which include galaxies via HOD modeling, and within the TNG300 suite of hydrodynamic simulations.


For each of the seven input parameters, ten realizations of the model were produced. Once the catalogs were generated, the three IA correlation functions were estimated according to Equations \ref{eq:xir} - \ref{eq:etar}. This led to a data matrix of the shape $(110,526; 7; 100; 3; 20)$. In other words, we have $110,526$ tuples of seven input values that produced ten realizations of three correlation functions that each occupy twenty radial bins. This catalog is used to train the IA emulator (\texttt{IAEmu}) surrogate model introduced in \cite{pandya2024learning}. \texttt{IAEmu} is a neural network based emulator designed to predict galaxy IA statistics from a given HOD, with the alignments parameterized according to the two parameter family $\mu_{cen}$ and $\mu_{sat}$ \citep{van2024empirical}. These $110,526$ tuples are broken down in training, validation, and testing sets for IAEmu. This is summarized in Table \ref{tab:iatab}.

\begin{table}[!htb]
    \centering
    \begin{tabular}{cccccc}
        Training Set & Validation Set & Testing Set & Input Parameters & Output Bins & Realizations \\ \hline
         $77,368 ~(70\%)$ & $ 11,052 ~(10\%)$ & $22,105 ~(20\%)$ & $7$ & $20$ & $10$ \\
    \end{tabular}
    \caption{Data set statistics for the IA catalogs}
    \label{tab:iatab}
\end{table}

\section{Algorithm Overview} \label{sec:algo}

In this section, we outline our algorithms used to estimate correlations. Our algorithm has three variants, one geared toward estimating clustering uncertainties and the other two towards differentiability. Variations on the algorithm geared toward estimating clustering uncertainties are needed because estimating uncertainty requires sampling, which is incompatible with differentiation. There are a variety of ways to adapt sampling to be a differentiable process, which we explore in depth in \S\ref{sec:differentiability}. Throughout this work, we use most of the formalism laid out in \cite{bezdek1984fcm}. Our formalism differs in that it uses a distance function to compute angular distances on the sky rather than using a matrix to define a norm. We also define the quantity matrix, which allows us incorporate the quantities we wish to correlate into the original formalism. For details on the fuzzy sets formalism, see \cite{zadeh1965fuzzy} and \cite[page 38,][]{aluffi2021algebra}. Algorithm 1 first uses K-means$++$ \citep{kapoor2017comparative} to initialize a series of cluster centers. This algorithm adds initial cluster centers sequentially by considering the distances to centers that have already been determined. This initialization has been shown to help clustering algorithms and is better than choosing initial cluster centers at random \citep{kapoor2017comparative}. We next use the fuzzy-c-means algorithm \citep{bezdek1984fcm} to obtain final cluster centers and weight assignments that minimize the functional 

\begin{equation}
    J_{m}(U,v) = \sum_{k=1}^N \sum_{i=1}^c \left(U_{ik}\right)^m \lVert y_k - v_i \rVert ^2
\end{equation}
where \begin{itemize}
    \item $Y = \left\{y_1, \hdots, y_n\right\}$ are the object positions,
    \item $c$ is the number of clusters in $Y$; $2 \leq c < n$,
    \item $m$ is the weighting exponent (or fuzziness); $1 \leq m < \infty$,
    \item $U$ is the fuzzy c-partition of $Y$; $U \in \mathbb{R}^{c \times N}$
    \item $v = (v_1, \hdots v_c)$ is a vector of centers,
    \item $v_i = (v_{i1}, \hdots, v_{in})$, is the center of cluster $i$.
\end{itemize}

We also define $q_{ij}$ with $1 \leq i \leq M$ and $1 \leq j \leq N$ as the quantity matrix. Each row represents a quantity of interest that we wish to correlate and each column represents a different object. In this work, there will be two rows corresponding to two shears. Our formalism differs from \cite{bezdek1984fcm} in that the norm on $\lVert y_k - v_i \rVert ^2$ is not induced by a matrix. Instead, we use the Vincenty formula given by Equation \ref{eq:vincenty} to compute angular separations on the sky between two angular coordinates $(\phi_1, \lambda_1), (\phi_2, \lambda_2)$:

\begin{equation}
    \Delta \sigma = \arctan \left( \sqrt{\left(\cos \phi_2 \cdot \sin \Delta \lambda\right)^2 + \left(\cos \phi_1 \cdot \sin \phi_2 - \sin \phi_1 \cdot \cos \phi_2 \cdot \cos \Delta \lambda \right)^2 }, \sin \phi_1 \cdot \sin \phi_2 + \cos \phi_1 \cdot \cos \phi_2 \cdot \cos \Delta \lambda \right).
    \label{eq:vincenty}
\end{equation}

While this does not have a compact matrix representation, it is stable at all numerical scales \citep{vincenty1975direct}.  In this way, we are sacrificing speed for stability. This design choice is discussed more in $\S$\ref{sec:differentiability}. The two argument $\arctan$ function is used to constrain the solution to the correct quadrant. Our usage of the Vincenty formula also assures that our distance estimates are consistent with Astropy \citep{robitaille2013astropy}, which also uses the Vincenty formula for this task.

Our implementation of fuzzy-c-means works as follows:
\begin{enumerate}
\item Initialize centers via K-means$++$ and weights via random initialization.
    \item Update centers via \begin{equation}
        \hat{v}_i = \frac{\sum_{k=1}^N \left(\hat{U}_{ik}\right)^m y_k}{\sum_{k=1}^N \left(\hat{U}_{ik}\right)^m} ; 1 \leq i \leq c
    \end{equation}
    \item Update weights via \begin{equation}
        \hat{U}_{ik} = \left( \sum_{j=1}^c \left(\frac{\hat{d}_{ik}}{\hat{d}_{jk}}\right)^{\frac{2}{m-1}} \right)^{-1}
    \end{equation} where the distances $d$ are given by Equation \ref{eq:vincenty}.
    \item Repeat steps 2 and 3 until either the maximum number of iterations is reached or the difference in weights between iterations is less than a user-provided tolerance and matrix norm.
\end{enumerate}

After obtaining final centers $v$ and weights $U$, we can assign each galaxy to a cluster center via sampling. Each row in $U^T$ represents the normalized probability distribution that an object belongs to a cluster with center $v_i$. Thus, each object is assigned to a cluster via 
\begin{equation}
    M_i \sim \text{Categorical}(U^T_i). \label{eq:sample}
\end{equation}
 
We define the quantities of each cluster as the average of all the quantities of the objects assigned to it. Note that this is unnecessary if there are no quantities. This may arise if one is computing a position-position autocorrelation or similar. From here, we can proceed by computing all of the pairwise distances between clusters and then use a known estimator to compute the correlation in each distance bin. The whole of Algorithm \ref{alg:probabilistic_estimation} is recapitulated in the algorithm block below.

\begin{algorithm}[H]
\large
\caption{Probabilistic Estimation} \label{alg:probabilistic_estimation}
\begin{algorithmic}[1]
       \STATE Initialize centers $v$ using K-Means$++$
    \STATE Compute new centers $v$ and weights $U$ using fuzzy-c-means
    \STATE Assign objects to clusters via $M_i \sim \text{Categorical}(U_i)$
    \STATE Define new objects with centers $v$ and quantities given by the average of all objects in the cluster
    \STATE Compute all pairwise distances $d$ between objects using Equation \ref{eq:vincenty}
    \STATE Bin the pairwise distances by spatial separation
    \FOR{each bin $b_i$}
        \STATE Evaluate correlation estimator on binned object distance pairs 
    \ENDFOR
\end{algorithmic}
\end{algorithm}

\begin{figure}[b]
    \centering
    \includegraphics[width=0.75\linewidth]{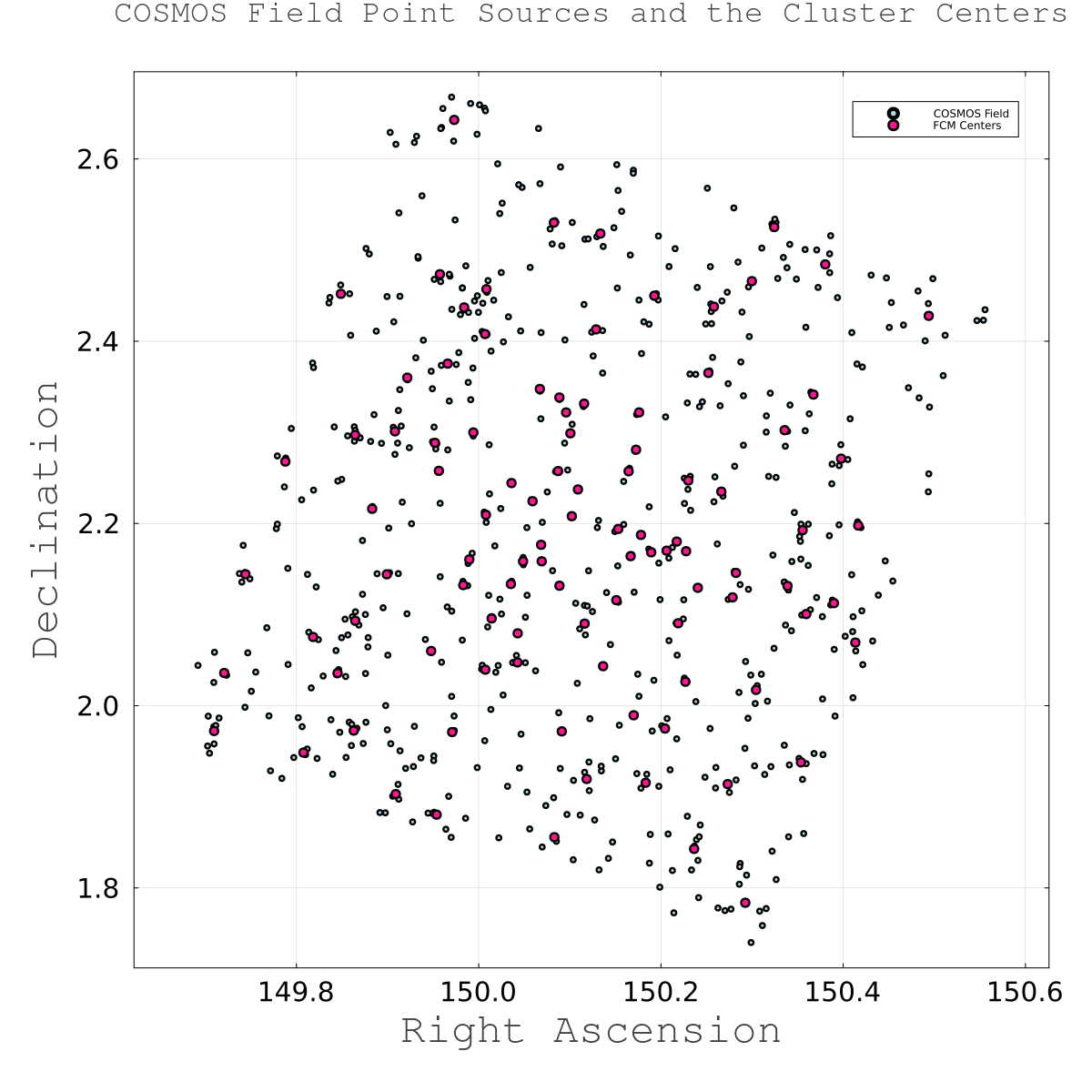}
    \caption{The cluster centroids found via fuzzy-c-means overlayed on the point sources of the underlying COSMOS field}
    \label{fig:scatter}
\end{figure}

\begin{figure}[!htb]
    \centering
    \vspace{5pt}
    \includegraphics[width=0.85\linewidth]{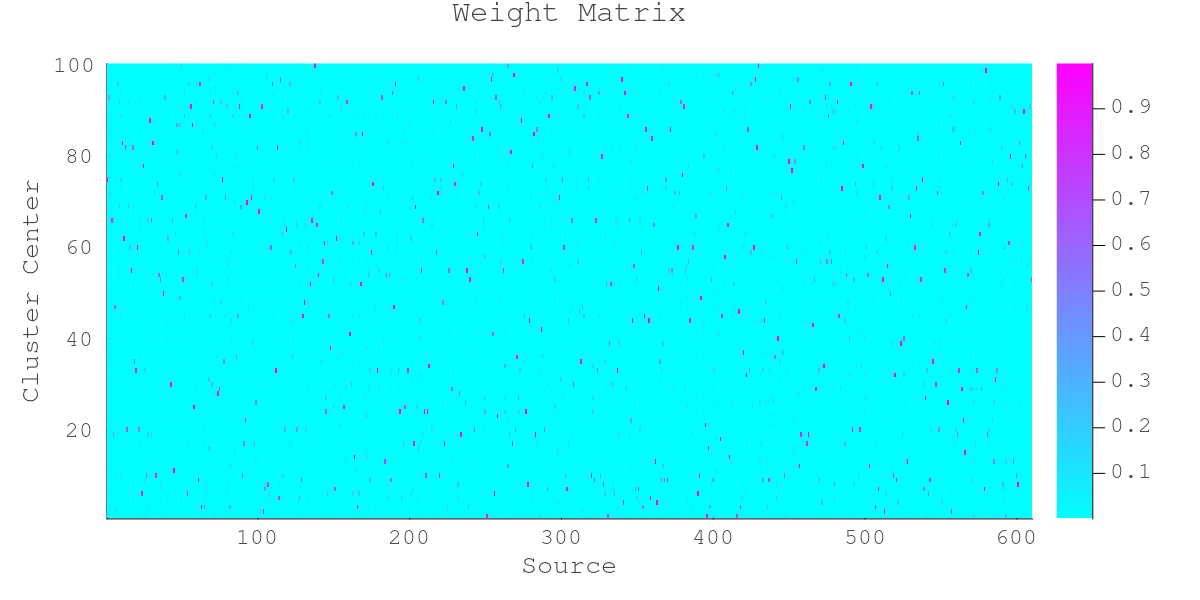}
    \caption{An example weight matrix $U$ produced via fuzzy-c-means for $610$ point sources in the COSMOS field and $100$ cluster centers.}
    \label{fig:weight}
\end{figure}

Algorithm \ref{alg:differentiableestimator} is a differentiable analog to Algorithm \ref{alg:probabilistic_estimation}. Algorithm \ref{alg:probabilistic_estimation} goes from soft bin assignments to hard assignments by sampling the rows of $U$. By soft assignments, we mean that for each object the assignment probabilities are spread across the different clusters. This is in contrast to our definition of a hard assignment in \S\ref{sec:intro}. This transition from hard assignment to soft assignments is the first obstacle to be amended in Algorithm \ref{alg:differentiableestimator}, since sampling is not a differentiable process. There are at least three ways to adapt this step in Algorithm \ref{alg:probabilistic_estimation} to become differentiable, which we will hereafter refer to as approaches a, b, and c: \begin{enumerate}[label=(\alph*)]
    \item Use the Gumbel Max reparameterization trick to enable differentiation through the sampling process \citep{maddison2016concrete, jang2016categorical} \footnote{An example implementation of this can be found here: \url{https://github.com/cassanof/gumbel-bucket-rs}.}.
    \item Avoid sampling entirely and instead compute weighted averages to get the quantities for each cluster. The quantities are weighted by how likely they are to belong to an object in a given cluster and then averaged. This is done for each cluster.
    \item Assign each object to a cluster with the highest membership probability via \textsc{argmax} (effectively proceeding with hard assignments). Adapt the membership matrix such that there is a $1$ in the entries $U_{ij}$ where object $j$ most likely belongs to cluster $i$. Then, make new objects by averaging the quantities in each cluster. During the gradient calculation, approximate the gradients for each object as the gradient of the new cluster object it belongs to. In other words, if object $i$ is in cluster $j$, approximate the gradient of a correlation function with respect to object $i$ as the gradient of the correlation function with respect to the new object generated from cluster $j$.
\end{enumerate}  We explore approaches $b$ and $c$ because they are simple to compute via matrix multiplication. For these approaches to work, we must also normalize the rows of $U$. We define 
\begin{equation}
    \Tilde{U}_{ij} = \frac{U^T_{ij}}{\sum_{k=1}^m U^T_{kj}}.
\end{equation} Now we can express our weighted average with the matrix product 
\begin{equation}
    \Tilde{q} = q\Tilde{U}.
\end{equation} We see that $\Tilde{q} \in \mathbb{R}^{M \times c}$, representing the $M$ quantities for each cluster $c$. Typical correlations choose $M$ to be $2$ or $3$. For example, if we were computing the $\xi_+$ shear-shear correlation, we would have two quantities $e$ and $e^*$ for each cluster.

Another adaptation we need to make from algorithm \ref{alg:probabilistic_estimation} is the hard assignment of distance pairs to bins. Algorithm \ref{alg:probabilistic_estimation} takes all of the distance pairs that fall into a distance bin and runs an estimator to get the correlation value for that bin. The differentiable analog is to compute the estimator over \textit{all} distance pairs, where the contribution of each distance pair is weighted by how likely it is to fall into a given bin with bin edges $a$ and $b$. In the case of the shear-shear correlation, this extends naturally the known estimator in Equation \ref{eq:estimator}. We determine the weights $w_i w_j$ using the product of two sigmoid functions, which gives high importance to distances within the bin and almost no importance to objects that are outside of it. We explore two weighting functions. The first is given by 
\begin{equation}
    W(a, x, b) = \frac{1}{1 + e^{-\alpha(x-a)}} \cdot \frac{1}{1 + e^{\alpha(x-b)}} \label{eq:sigmoids}
\end{equation}
and the second by
\begin{equation}
    W(a, x, b) = e^{-\alpha\left(x - \frac{b-a}{2}\right)^2}. \label{eq:gaussian_w}
\end{equation}

\begin{figure}[!htb]
    \centering
    \includegraphics[width=0.65\linewidth]{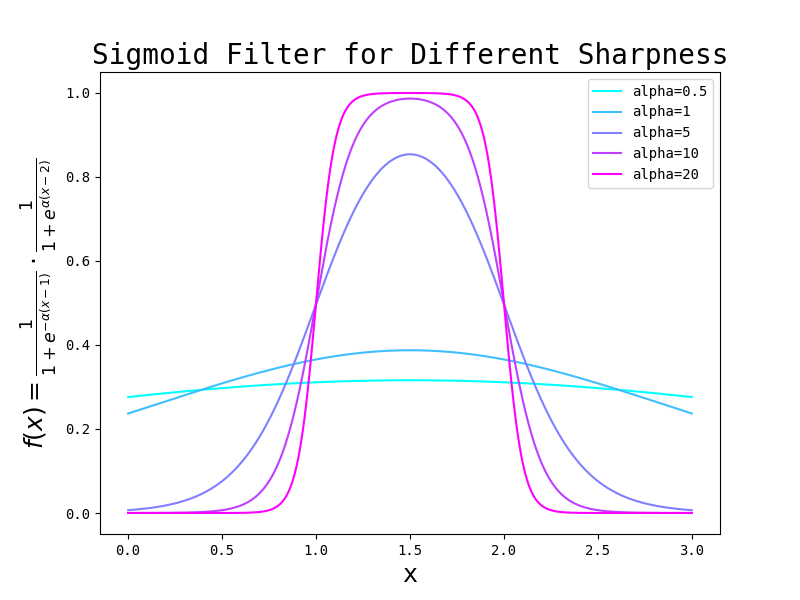}
    \caption{An example weighting for estimates for a distance bin bounded below by $1$ and above by $2$ for various sharpnesses.}
    \label{fig:sigmoids}
\end{figure}

Both weighting functions exploit the sharp decline of the exponential function to quickly decrease the weights. An example of this for the first weighting function is shown in Figure \ref{fig:sigmoids} for distances between $1$ and $2$. Algorithm \ref{alg:differentiableestimator} is summarized in the algorithm block below. Given that both weighting functions can have their decay controlled by the $\alpha$ parameter, we found that both weighting functions can be tuned to produce correlation outputs consistent with other packages used for estimating correlation functions.

\begin{algorithm}[H]
\large
\caption{Differentiable Estimation} \label{alg:differentiableestimator}
\begin{algorithmic}[1]
       \STATE Initialize centers $v$ using K-Means$++$
    \STATE Compute new centers $v$ and weights $U$ using fuzzy-c-means
    \STATE Define new objects with centers $v$ and quantities $q\Tilde{U}$
    \STATE Compute all pairwise distances $d$ between objects using Equation \ref{eq:vincenty}
    \STATE Compute the probability that each object pair falls in distance bin $b_i$ by weighting by the sigmoid function in Equation \ref{eq:sigmoids} or the Gaussian function in Equation \ref{eq:gaussian_w}
    \FOR{each bin $b_i$}
        \STATE Evaluate correlation estimator on galaxy pairs weighted by their probability of being in bin $b_i$
    \ENDFOR
\end{algorithmic}
\end{algorithm}

As we will see, there are some computational hurdles that any differentiable estimator will inevitably face that can make differentiability intractable in practice. These can be circumnavigated using a surrogate model, as outlined in Algorithm~\ref{alg:surrogatealg}.

\begin{algorithm}[H]
\large
\caption{Surrogate Differentiable Estimation}
\label{alg:surrogatealg}
\begin{algorithmic}[1]
       \STATE Simulate realizations of a correlation function with astrophysical model parameters $\phi$
    \STATE Divide correlation function outputs into training, validation, and testing sets
    \STATE Train a model with learnable parameters $\theta$ to learn the relationship between astrophysical model parameters and correlation functions from a (not necessarily differentiable) forward model, $\xi_\theta(\phi)$
    \STATE Once trained, perform inference on parameters of choice
    \STATE Optionally optimize via gradient descent the parameters $\phi$ that minimize some loss function $L(\xi_\theta(\phi))$
    \STATE Optionally retrieve via HMC the parameter posteriors $p(\phi | \text{data})$ with priors $p(\phi)$ and a likelihood $p(\text{data}|\phi)$
\end{algorithmic}
\end{algorithm}

\section{Model Uncertainty} \label{sec:modeluq}

In this section, we discuss the results of Algorithm \ref{alg:probabilistic_estimation} on our point source catalog dataset, which we compare with a bootstrapping baseline to obtain epistemic and aleatoric uncertainties.

Figure \ref{fig:scatter} shows the cluster centers overlayed with the objects on the sky. The result indicates that the cluster centers reasonably approximate the spatial variation of the sources in the underlying COSMOS field. Figure \ref{fig:weight} shows the weight matrix after the objects are clustered with fuzzy-c-means.

\begin{figure}[!htb]
    \centering
    \includegraphics[width=0.95\linewidth]{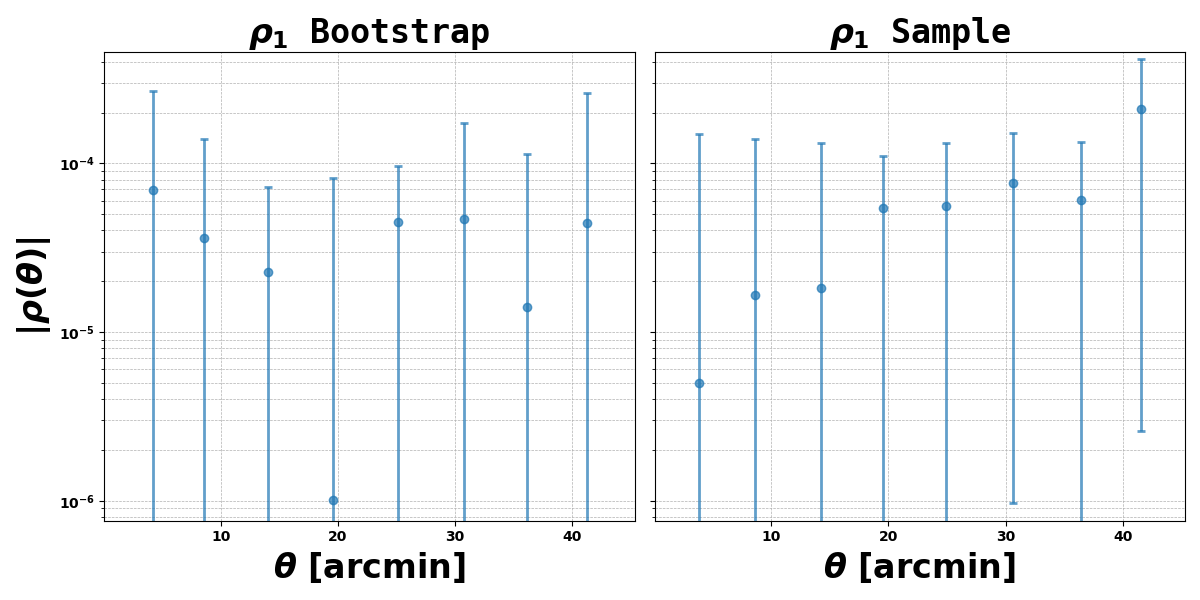}
    \caption{$\rho_1$ correlation function with data taken from the COSMOS-Web point source catalogs. Error bars drawn from bootstrap (left) and sampling (right) approaches. Negative correlations are shown in absolute value, and correlations are plotted in logarithmic scale.}
    \label{fig:rho1comp}
\end{figure}

\begin{figure}[!htb]
    \centering
    \includegraphics[width=0.9\linewidth]{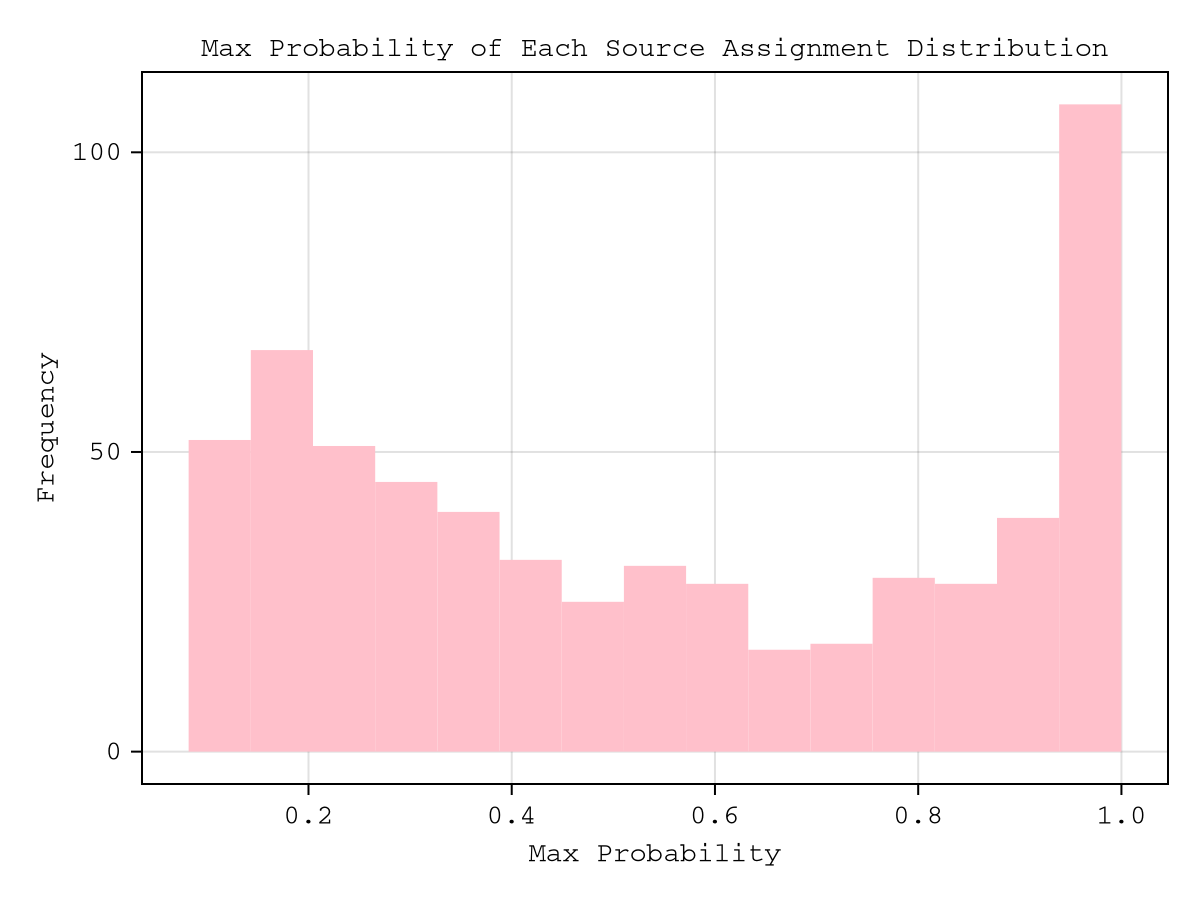}
    \includegraphics[width=0.9\linewidth]{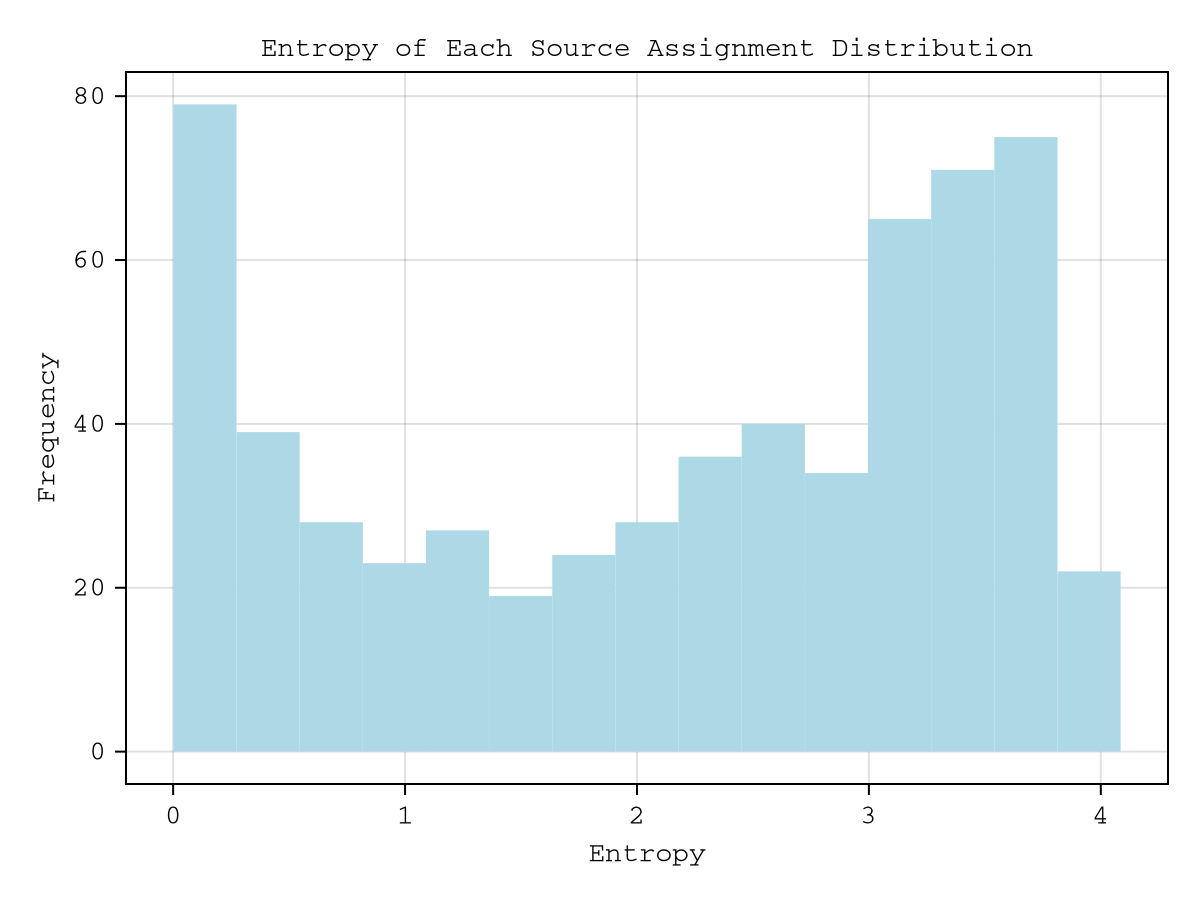}
    \caption{The maximum probability and entropy of the source assignment distributions. Each column in $U$ corresponds to a probability distribution. For each of those distributions, we compute the maximum probability and Shannon entropy.}
    \label{fig:distributions}
\end{figure}

Figure \ref{fig:distributions} shows the Shannon information entropy and maximum probability of each row in the weight matrix. While the majority of objects have a clear preference toward being assigned to a specific cluster, there is a non-trivial number of objects that could be reasonably assigned to multiple clusters. This indicates that the model is often not sure where the best assignment of an object to a cluster is. For this reason, we should expect the error bar associated with model uncertainty to be comparable to data uncertainty. This bears out in Figure \ref{fig:rho1comp} and Figures \ref{fig:rho2comp} - \ref{fig:rho5comp}.

We note that increasing or decreasing the number of clusters did not significantly reduce the size of the model uncertainty. This indicates that the source density in our COSMOS-Web data set is not large enough to beat down the errors associated with clustering. Recall that in \S\ref{sec:notation} we said that we are treating uncertainties as additive. Figures \ref{fig:rho1comp} and \ref{fig:rho2comp} - \ref{fig:rho5comp} therefore indicate that in many cases the epistemic uncertainties doubles the uncertainty compared to bootstrapping alone. This is especially true for the $\rho_1$, $\rho_2$, and $\rho_4$ statistics. For this reason, we suggest a ``na\"{\i}ve''\footnote{The use of the word na\"{\i}ve is in reference to its usage in \cite{jarvis2004skewness}.} approach to computing $\rho$ statistics, wherein no clusters are used and the correlation estimators are used directly on the catalogs. Given that we have on the order of $\sim 500$ objects, the number of pairwise distances is not too large to store and the number of distance calculations to compute is not prohibitively expensive.

\section{Differentiable Forward Model} \label{sec:differentiability}

In this section we outline the utility of algorithm \ref{alg:differentiableestimator} for gradient-based optimization. To do this, we create a simulated catalog of galaxies moving in a gravitational field. We time evolve the objects until some terminus $T_{\rm max}$, which was found by considering an object undergoing uniform circular motion and calculating how long it would take to complete one orbit. For simplicity, we analyze correlations across three linearly spaced angular separation bins. A max separation of $160$ arcmins is chosen to ensure each of the three bins have enough samples. We seek to differentiate the shear-shear correlation $\xi_+(\theta)$ with respect to the gravitational constant $G$ at each angular separation bin $\theta_i$,
illustrating an example of enabling differentiability of a correlation with respect to an astrophysical model parameter. The ellipticities are predetermined, but the final positions are naturally dependent on the underlying physics. Since the catalog is produced by solving a series of ordinary differential equations (ODEs), computing the derivative just amounts to using the chain rule $\frac{\partial \xi}{\partial G} = \sum _{i=1}^N\frac{\partial \xi}{\partial \alpha_i } \frac{\partial \alpha_i}{\partial G} + \frac{\partial \xi}{\partial \delta_i} \frac{\partial \delta_i}{\partial G}$. Here, $\alpha$ denotes the right ascension and $\delta$ denotes the declination coordinates. We use the ODE solvers and automatic differentiation tools implemented in Julia \citep{bezanson2017julia}. Specifically, we use the OrdinaryDifferentialEquations.jl \citep{rackauckas2017adaptive} library for working with differential equations. We also tested both reverse mode automatic differentiation with Zygote.jl \citep{innes2019differentiable} and forward mode automatic differentiation with ForwardDiff.jl \citep{revels2016forward}. The latter proved to be more effective. We also built an implementation with Jax \citep{jax2018github} and Diffrax \citep{kidger2021on}, which is available on our GitHub repository. Our adoption of these tools is motivated by the discussion of \cite{berman2024state}.

\begin{figure}[!htb]
    \centering
    \includegraphics[width=0.75\linewidth]{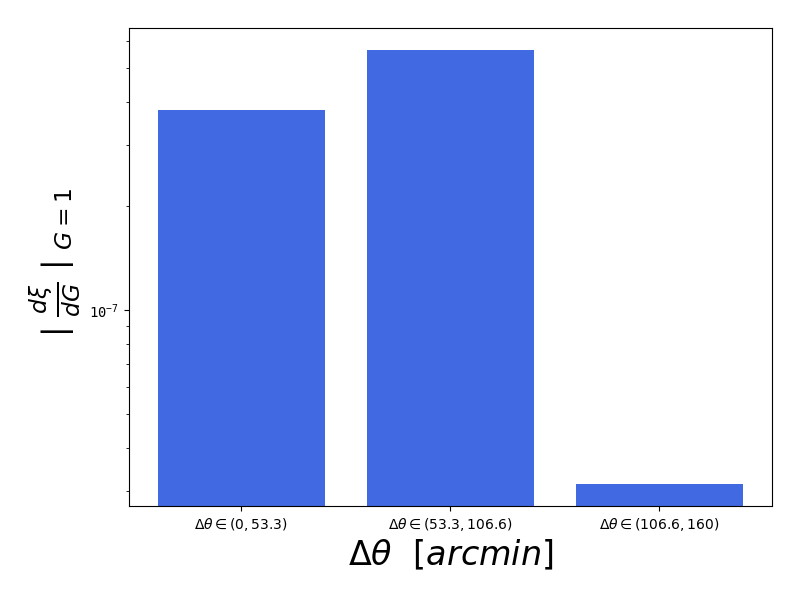}
    \caption{The gradient of the shear-shear correlation function $\xi$ with respect to the input model parameter $G$ at $G=1$.}
    \label{fig:dxidg}
\end{figure}

For both approaches b and c (as defined in $\S$\ref{sec:algo}), the naive approach to computing the gradient $\frac{\partial \xi}{\partial \alpha_i}$ or $\frac{\partial \xi}{\partial \delta_i}$ would involve tracking the computation graph through the entire fuzzy-c-means algorithm, which can sometimes take over $100$ iterations. Another complication is that the distance calculations and normalization computations cannot be expressed through matrix multiplication, making it difficult to accelerate the gradient calculations with a GPU. For approach $b$, taking the gradients in this manner proved to be prohibitively expensive via reverse mode automatic differentiation, for the reasons outlined above. With forward mode differentiation, the gradient calculation was on the order of a minute. This is expected, since we are differentiating several output bins with respect to $1$ input (gravity).

One workaround to the computation time bottleneck is to pre-compute the final centers and weights with fuzzy-c-means, then update the weights and centers for only a single iteration. This makes the final clustering a function of the data without keeping the entire history of the clustering procedure in the computation graph. Using this ``trick'' to approximate the true gradient, we can roughly halve the computation time needed to solve for the derivative $\frac{\partial \xi}{\partial G}$. However, when using approach $b$, issues related to precision still exist. When using a large number of clusters, we found that the Jacobian relating the dependence of weights on the input data was sparse. The sparsity result is expected, as local perturbations of input positions, by design, preferentially impact the assignment probabilities of nearby cluster centers. As a result, we are left with vanishing gradients, and the algorithm becomes unstable as we scale the number of clusters, producing \texttt{NaN} outputs. We note that these \texttt{NaN} issues did not occur on randomly computed inputs used for testing the differentiability of each individual function, indicating that the reliability of this method will vary for problems of different conditioning. This also confirms that while each individual step in the algorithm is differentiable, instabilities can still occur from gradient multiplication. It is worth noting that multiplying a matrix with another matrix that is sparse does not mean the resulting matrix is necessarily sparse. Thus, one Jacobian \textit{alone} being sparse is not the problem, rather, its the effect of a sparse Jacobian when propagated through the chain rule. Because we are reasoning about gradients, we want to ensure our representations are dense. We note that this is in contrast to other numerical problems where sparsity is necessary to maintain computational efficiency when dealing with large matrices \citep[chapter 10 of][]{saad2003iterative, rosen2022accelerating}.

Approach $c$ is able to circumnavigate some of these issues. As evidenced by Figure \ref{fig:dxidg}, we are successfully able to differentiate the outputs of our correlation function in each distance bin with respect to the input astrophysical model parameter $G$ at the point $G = 1$. However, issues with scaling the number of data points still persist, as the computation becomes too large to run on a CPU well before $500$ galaxies are simulated via the forward model.

The precision and computation time issues suggest that the simple solution is not always sufficient for enabling the differentiability of correlation functions with respect to the underlying model inputs. Surrogate models, however, have demonstrated the ability to succesfully emulate correlation function estimators \citep{van2024empirical, pandya2024learning, pandya2025iaemulearninggalaxyintrinsic}. These surrogate models can be as simple as linear regression, or more involved functions such as those represented by a neural network (NN). NNs, by definition, are differentiable models (i.e. trained with gradient descent) and enjoy GPU accelerated computations.

\section{Differentiable Surrogates} \label{sec:surrogates}

In this section, we show how Algorithm \ref{alg:surrogatealg} can be used to differentiate through a correlation function and present a compelling science case that supports its utility. In particular, we utilize the surrogate model developed in \cite{pandya2024learning, pandya2025iaemulearninggalaxyintrinsic}, \texttt{IAEmu}, which estimates galaxy IA correlations from an underlying set of HOD parameters. We use \texttt{IAEmu} to optimize model parameters that minimize the IA correlations, to study regions of parameter space that correspond to minimal IA contamination and are thus of physical interest. \texttt{IAEmu} serves as an emulator for Halotools-IA, which extends traditional HOD modeling to include IA information through a two parameter family, $\mu_\text{cen}$ and $\mu_\text{sat}$, which govern central and satellite alignment strengths, respectively. More specifically, \texttt{IAEmu} outputs the galaxy position-position ($\xi$), position-orientation ($\omega$), and orientation-orientation ($\eta$) correlations as well as estimates of the aleatoric uncertainty (shape noise) of each correlation. This noise is not physical, and corresponds to variance across realizations of the underlying HOD due to limited cosmological volumes when conducting the HOD simulations.

We pose this optimization problem because IA often act as a contaminant for weak lensing measurements, and moreover, there are known regions of parameter space that lead to low correlations. Thus, with this experiment, we are able to demonstrate that a differentiable forward model allows one to learn something about the underlying physics of IA. In other words, we are able to recover a known parameter space even without hard-coding the physical laws into the surrogate directly. This is distinct from \cite{pandya2025iaemulearninggalaxyintrinsic}, who use the emulator to find the IA parameters that best explain a set of observations rather than the parameters that result in the minimized correlation function. \footnote{The exact correlation that would be subtracted is the gravitationally lensed shape - intrinsic shape $(GI)$ correlation \citep{lamman2023ia}; however, the strength of that correlation is closely related to that of $\omega$.} 

\begin{figure}[!htb]
    \centering
    \includegraphics[width=1\linewidth]{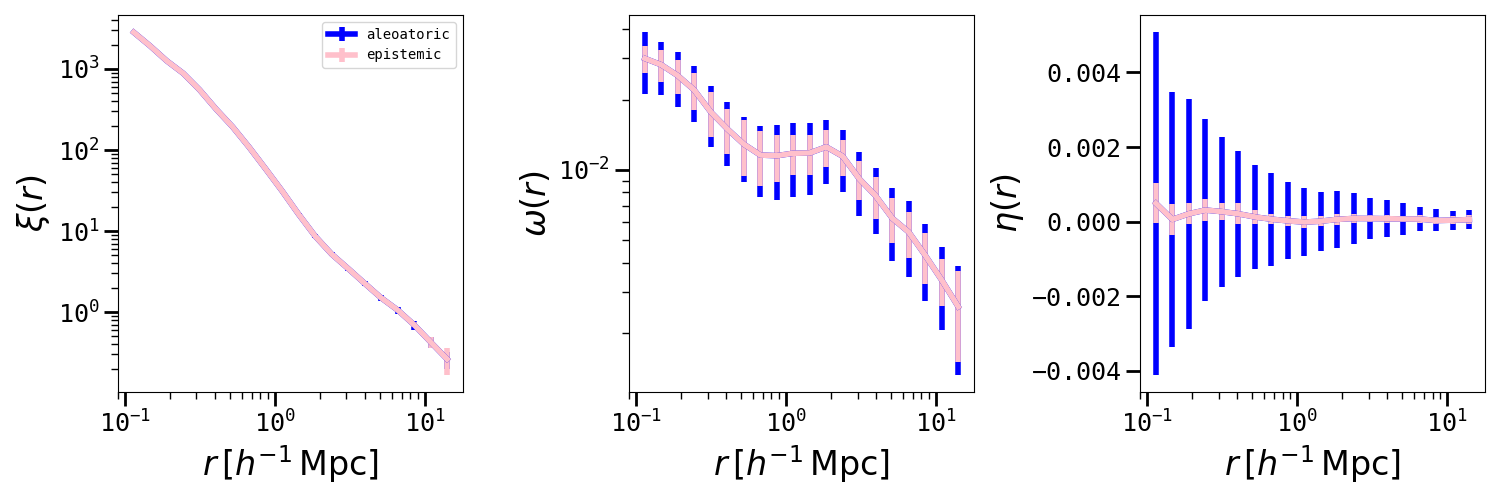}
    \caption{IAEmu predictions for a given initialization of the seven halo-based modeling parameters, including $\xi(r)$ on the left, $\omega(r)$ in the middle, and $\eta(r)$  on the right. For each of the twenty radial bins, the value of the correlation, as well as aleatoric and epistemic uncertainties, are provided. Log-log scaling is used to better see the dynamic range of the correlations and radial dependence.}
    \label{fig:ia_corr_default}
\end{figure}

The surrogate model introduced in \cite{pandya2024learning, pandya2025iaemulearninggalaxyintrinsic} is a deep neural network with a multilayer perceptron encoder and three convolutional neural network decoder heads for each of the three IA correlation functions. The model is also constructed to predict aleatoric uncertainties per-bin for each correlation, and epistemic uncertainties via the Monte Carlo dropout technique \citep{hullermeier2021aleatoric}. For a single initialization of the seven Halotools-IA parameters, Figure \ref{fig:ia_corr_default} shows all three IA correlation functions. Our study focuses exclusively on the $\omega(r)$ correlation. This is because $\xi(r)$ is akin to galaxy clustering and therefore does not contain any IA information itself. In addition, the galaxy shape noise that is present in $\eta(r)$ significantly increases the difficulty of extracting relevant signal for gradient-based optimization pipelines, as seen in Figure \ref{fig:ia_corr_default}.
This leaves us with $\omega(r)$, as it contains IA information but is significantly less obscured by shape noise when compared to $\eta(r)$. 
This is seen quantitatively in Figure \ref{fig:jacobian}, which shows the Jacobian of each correlation as a function of the model inputs. It is seen that the gradient magnitudes of \texttt{IAEmu} are in general larger than that of $\eta$. These gradients are also stable from bin-to-bin, and can further be accessed in less than a second. While there may be some disagreement with the exact sensitivity that the Halotools-IA\footnote{\url{https://github.com/astropy/halotools}} model presented in \cite{van2024empirical} would predict, we will see that the gradients information from \texttt{IAEmu} further affirms that it is an accurate emulator for Halootols-IA.

To isolate regions of parameter space in $\mu_\text{cen}$ and $\mu_\text{sat}$ that minimize the IA correlation $\omega(r)$, we utilize Hamiltonian Monte Carlo (HMC). HMC is a gradient based Monte Carlo technique which leverages the differentiability of NNs and uses gradient information to efficiently traverse the input parameter space towards regions that maximize the posterior probability. For a general review of HMC, see \cite{betancourt2017conceptual}. Our HMC experiment mirrors \cite{pandya2025iaemulearninggalaxyintrinsic}. We run our HMC with a No U-Turn Sampler (NUTS) \citep{hoffman2014no} with $1000$ warmup steps and $1000$ samples. We fix the HOD parameters according to the fiducial values present in IllustrisTNG300-1 \citep{nelson2019illustristng} and Table C1 of \cite{van2024empirical}. This gives us three sets of HOD parameters, which each correspond to a different stellar mass cutoff in galaxies.  We impose uniform priors of $[-1, 1]$ on $\mu_{\rm cen}$ and $\mu_{\rm sat}$ and use a multivariate normal distribution centered at zero as our likelihood, with a covariance given by the IAEmu aleatoric uncertainty prediction. In this way, we encourage HMC to arrive at regions of parameter space that minimize the $\omega$ correlation, while also incorporating the inherent shape noise into the likelihood. 

\begin{figure*}[!htb]
    \centering
    \subfigure[$\log M_* \geq 10.0$]{
        \includegraphics[width=0.31\textwidth]{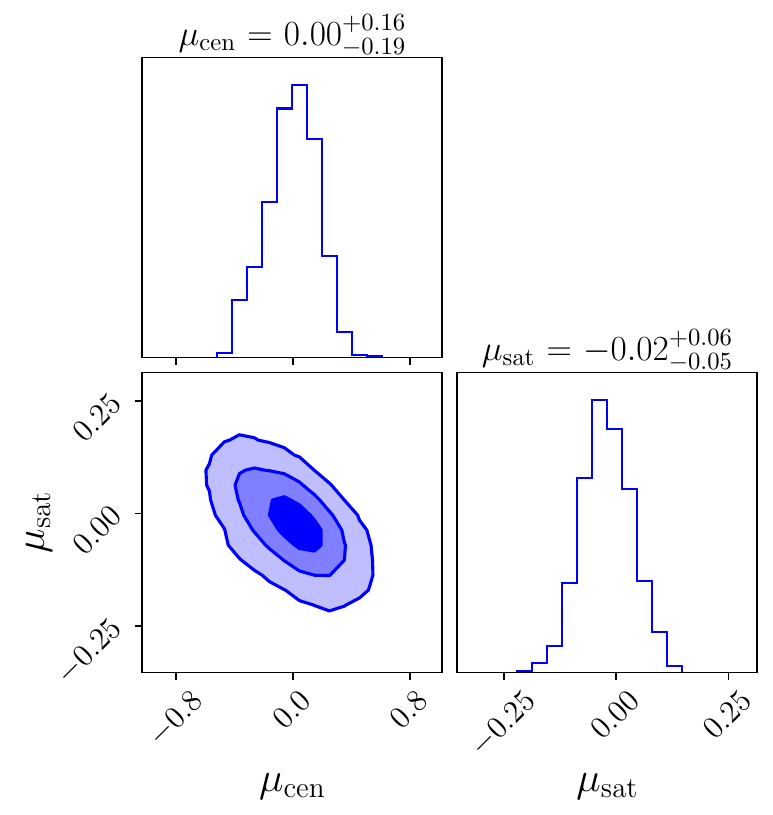}
        \label{fig:figure1}
    }
    \subfigure[$\log M_* \geq 9.5$]{
        \includegraphics[width=0.31\textwidth]{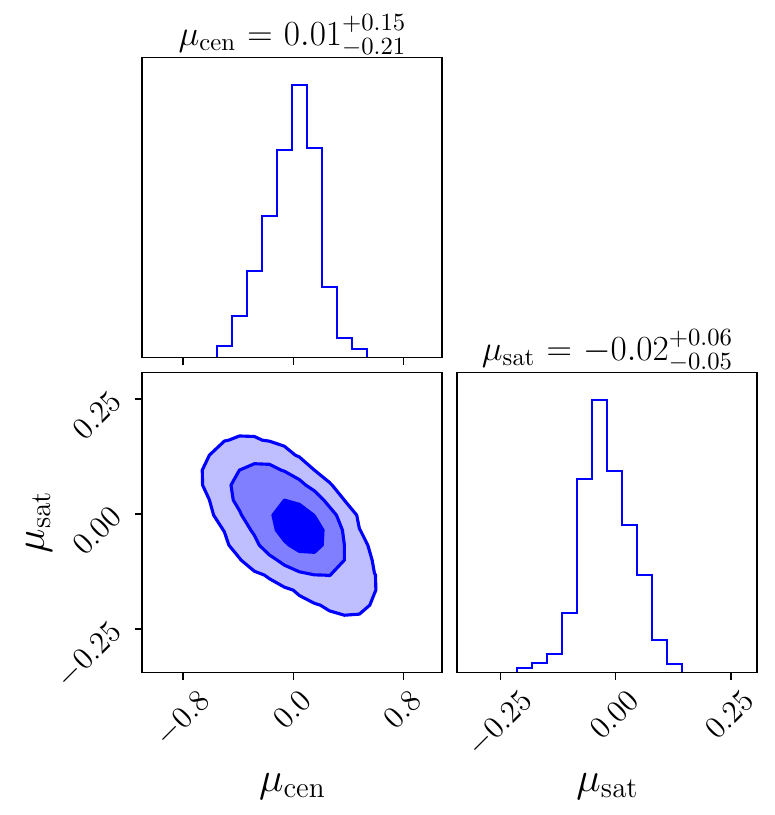}
        \label{fig:figure2}
    }
    \subfigure[$\log M_* \geq 9.0$]{
        \includegraphics[width=0.31\textwidth]{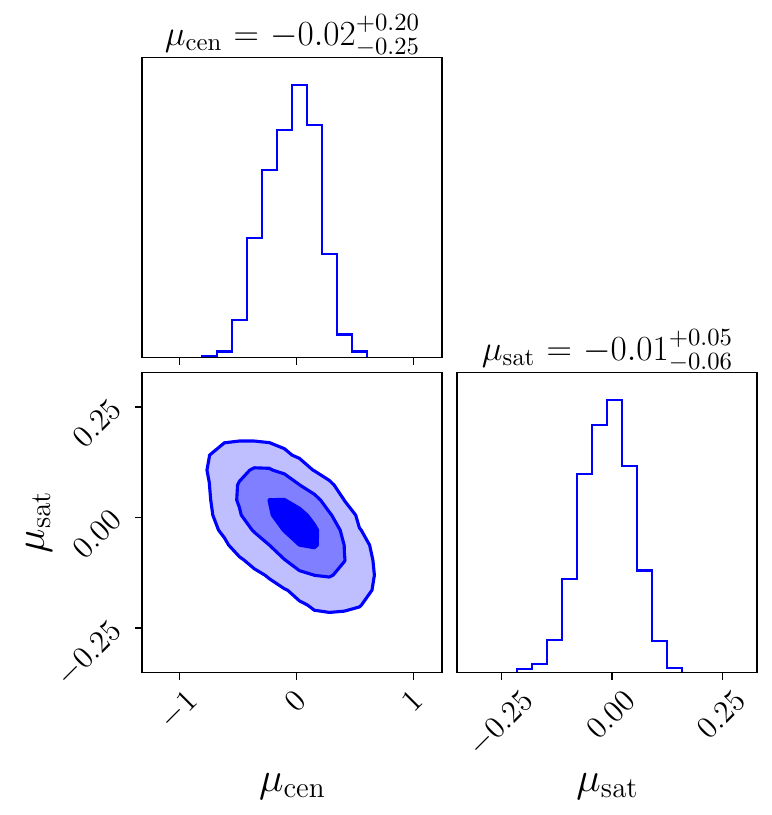}
        \label{fig:figure3}
    }
    \caption{Posterior constraints on $\mu_{cen}$ and $\mu_{sat}$ obtained via HMC for three different sets of HOD parameters. Each set roughly corresponds to a different stellar mass cutoff in galaxies. We show results for $\log M_* \geq 10.0$ on the left, $\log M_* \geq 9.5$ in the middle, and $\log M_* \geq 9.0$ on the right. Contours represent $1\sigma$, $2\sigma$, and $3\sigma$ confidence intervals.}
    \label{fig:corner}
\end{figure*}

\begin{figure*}[!htb]
    \centering
    \subfigure[$\log M_* \geq 10.0$]{
        \includegraphics[width=0.31\textwidth]{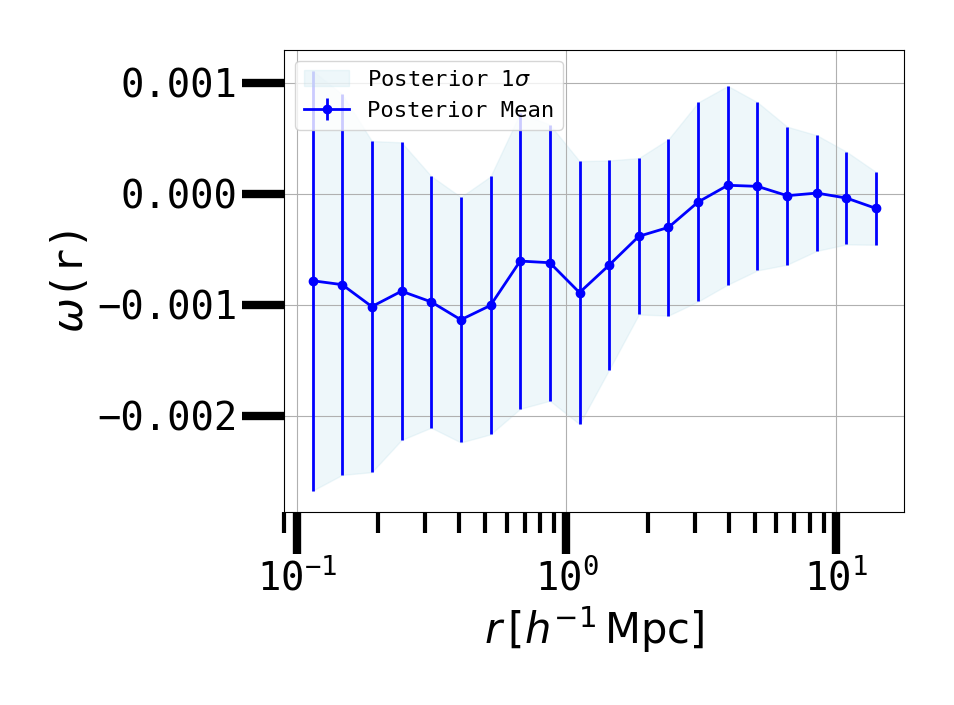}
        \label{fig:hmccor3}
    }
    \subfigure[$\log M_* \geq 9.5$]{
        \includegraphics[width=0.31\textwidth]{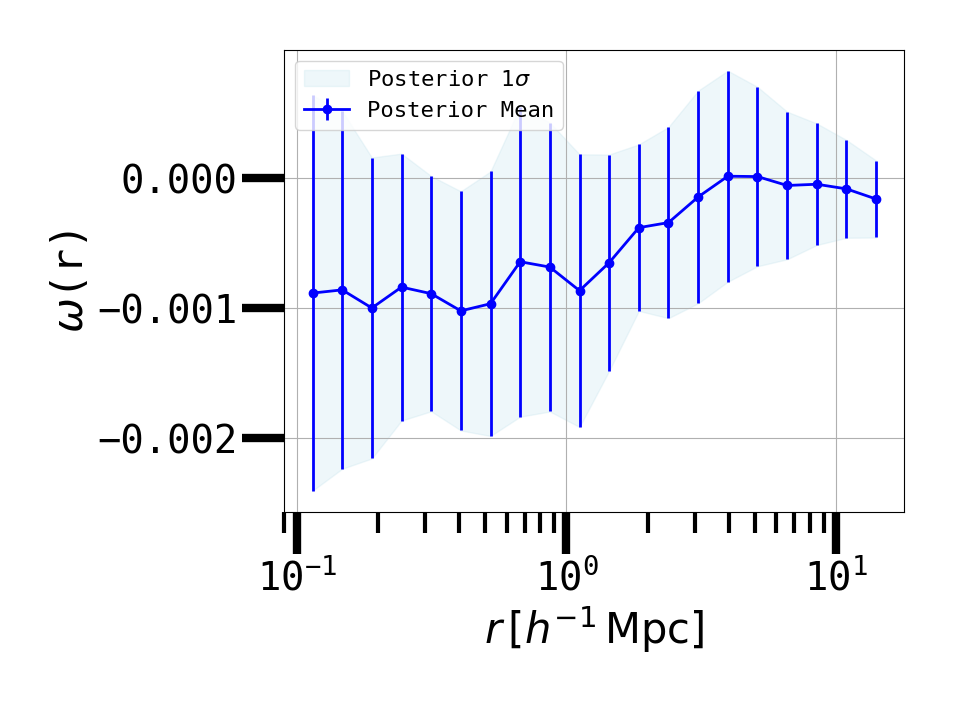}
        \label{fig:hmccor2}
    }
    \subfigure[$\log M_* \geq 9.0$]{
        \includegraphics[width=0.31\textwidth]{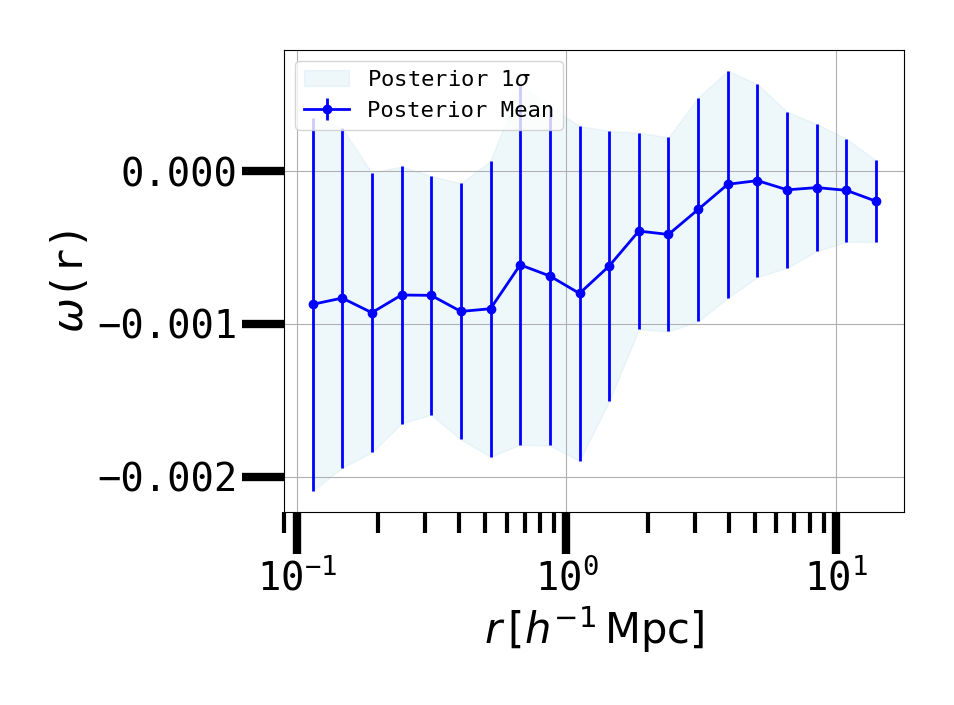}
        \label{fig:hmccor1}
    }
    \caption{Correlation functions sampled from the posterior distributions presented in \ref{fig:corner}. The solid represents the average correlation value in all bins from the $1000$ samples of $(\mu_{cen}, \mu_{sat})$ pairs obtained via HMC. The error bar represents the standard deviation of the correlation value for that same sample.}
    \label{fig:hmcpredict}
\end{figure*}

Figures \ref{fig:corner} and \ref{fig:hmcpredict} confirm that $\mu_\text{cen}$ and $\mu_\text{sat}$ values of zero minimize $\omega$, with some degeneracy as indicated by the diagonal shape of the posterior. While values of $\mu_{cen} = 0 = \mu_{sat}$ are likely to minimize the correlation function, regions where the signs of $\mu_{cen}$ and $\mu_{sat}$ are opposite can also result in low correlations. Physically, this corresponds to scenarios wherein the negative (i.e. perpendicular) alignments of satellite galaxies can oppose the positive (i.e. parallel) alignment of centrals, resulting a overall small value of $\omega(r)$.



\newpage
\section{Summary and Conclusions} \label{sec:conclusions}

In this work, we studied correlation functions along three axes: model uncertainty, differentiable forward models, and differentiable surrogates. To better understand the model uncertainty caused by clustering approximations, we used $\rho$ statistics as a case study. We made our assignment of objects to clusters probabilistic, and computed the mean and standard deviation of the resulting correlation estimates. Our analysis compared the uncertainty estimate to traditional bootstrapping methods. Our next experiment adapted our main algorithm to be automatically differentiable and applied this to the setting of gravitational simulations. Three different approaches of doing this were proposed, and of those, approach $c$ proved to be the most stable. Still, the calculation of the gradient was slow and approximate, motivating the use of surrogate methods. To that end, we ended by exploring surrogate solutions. We extended the work of \cite{pandya2024learning, pandya2025iaemulearninggalaxyintrinsic}, showing how we can exploit the gradients of our surrogate model to optimize over astrophysical model parameters. Specifically, we studied galaxy intrinsic alignment, and found which astrophysical model parameters were most responsible for high / low correlations (Figures \ref{fig:corner}-\ref{fig:hmcpredict}). 

On uncertainty, it was found that the process of clustering objects before computing correlations can cause measurable uncertainties in limited data settings, such as with the COSMOS-Web PSF modeling efforts. That is, the model uncertainty was just as severe as the data uncertanties found via bootstrapping. In these circumstances, we conclude that the na\"{\i}ve approach, which considers all distance pairs without clustering to be most appropriate. Moreover, we find that our method of probabilistic clustering allows us to address one of the fundamental concerns in uncertainty quantification, that being the distinction between epistemic and aleatoric uncertainties. Again, this is quantified in Figures \ref{fig:rho1comp}-\ref{fig:rho5comp}, where the error bars for epistemic and aleatoric uncertainties are shown for each of the five $\rho$ statistics. 

On differentiability, we showed that the simplest differentiable algorithm has features that makes the computation of the gradients slow and unstable. These findings were demonstrated through attempts to differentiate a correlation function with respect to a single parameter $G$ that determined the results of a gravitational simulation. Three approaches, a-c, were outlined: Approach $a$ relied on the Gumbel-Max trick, approach $b$ used the weight matrix to compute new quantities via weighted averages, and approach $c$ used an approximate method to calculate the gradient. Of these, approach $c$ proved to be the most effective at circumnavigating these deficiencies. The features that caused slowness and instability were the normalization and distance function calls and sparse Jacobians that propagated through the chain rule. The normalization and distance function calls posed challenges for fast automatic differentiation as they cannot be expressed simply with matrix multiplication; the normalization function calls in particular are notorious bottlenecks for automatic differentiation pipelines. We also saw that many of our function calls yielded sparse Jacobians. Since the chain rule amounts to multiplying successive Jacobians, this caused numerical instabilities in computing the final derivative. It is worth noting that multiplying one matrix by a sparse matrix does not necessarily imply that the resulting matrix is sparse. However, through testing the function calls individually, we found that the numerical instabilities happened only once the gradients were multiplied together, indicating that matrix sparsity patterns in any one Jacobians is the culprit for numerical instability. This also explained why approach $c$ was most successful, as approach $c$ avoided the function call that related weights to the input data when taking the gradient. Approach $c$ was successfully able to capture the gradient of a correlation function with respect to an input cosmology. While there remains future work in optimizing approach $c$ for GPU support and overall making it faster, it proved to be the most effective way to write down correlation function estimators that are automatically differentiable. In the main, the differentiability experiment suggested that a more judicious way to construct differentiable correlation functions may be to use a surrogate model. 

On surrogates, we found that surrogate models are a simple and effective way to optimize astrophysical model parameters that determine a correlation function. Using galaxy IA as an example, a correlation function was minimized across $20$ radial bins. Using HMC, we were able to recover a known posterior over the IA parameters that lead to low correlation function outputs. A key caveat is that the true correlation functions that act as a contaminant are the $GI$ and $gI$ correlations \citep[as defined in][]{lamman2023ia}, but the $\omega(r)$ correlation is closely related in how it represents the presence of IA.

Prior to our work, properties such as model uncertainty and differentiability have been understudied in the context of correlation function estimators. This work is intended as a first step toward addressing these points. Our study is intentionally generic --- survey systematics can vary tremendously and the correlations being studied will also vary by science case. Rather than focusing on a specific scientific inquiry, we propose general algorithms and analyze their utility across different domains. Our GitHub artifact outlines and implements these algorithms, serving as a blueprint for future large-scale surveys in each of the contexts laid out in this work --- model uncertainty, differentiability, and surrogates. 

\section*{acknowledgements}

E.B. thanks Quinn Arbolante, \href{https://jakegines.in/}{Jacob Ginesin}, \href{https://www.luisali.com/}{Luisa Li}, Eric Zelikman for helpful discussions. E.B. is supported by a College of Science Dean's Office Research grant. E.B. also thanks Gregory Shomo for helpful guidance with the discovery computing cluster. S.P. acknowledges support from the National Science Foundation under Cooperative Agreement PHY-2019786 (The NSF AI Institute for Artificial Intelligence and Fundamental Interactions, \url{https://iaifi.org}). Support for the COSMOS-Web survey was provided by NASA through grant JWST-GO-01727 and HST-AR-15802 awarded by the Space Telescope Science Institute, which is operated by the Association of Universities for Research in Astronomy, Inc., under NASA contract NAS 526555. This work was made possible by utilizing the CANDIDE cluster at the Institut d’Astrophysique de Paris, which was funded through grants from the PNCG, CNES, DIM-ACAV, and the Cosmic Dawn Center and maintained by Stephane Rouberol. Further support was provided by Research Computing at Northeastern University. 

\newpage

\bibliography{main}

\clearpage

\appendix

\section{PSF Modeling} \label{sec:psf_appendix}

This analysis uses ShOpt.jl for PSF modeling because it was shown to be accurate and computationally fast in \cite{berman2024efficientpsfmodelingshoptjl, Berman2024}. Since we are expanding our input catalog to the entire survey area of the COSMOS field instead of individual tiles as in \cite{berman2024efficientpsfmodelingshoptjl}, we also expect astrometric distortions to be more severe (cf. Figure 4 in \cite{berman2024efficientpsfmodelingshoptjl}). Since ShOpt works in astrometric coordinates and is compatible with PSF cutout sizes needed for NIRCam, ShOpt is well-suited for this task. In contrast, PSFex exhibits bias over large survey areas \citep{jarvis2021dark} and PIFF does not handle cutout sizes needed for characterizing the NIRCam PSF gracefully \citep{berman2024efficientpsfmodelingshoptjl}. Tools like WebbPSF \citep{perrin2014updated} were not designed for the mosaiced images we are working with, making their application to our data non trivial. Other tools like STARRED \citep{Michalewicz2023} or PSFr \cite[Birrer et al. in prep, ][]{Birrer2021} have been applied to JWST data and could potentially also be applied to this data set. See also \cite{feng2025exoplanet}.

\clearpage

\section{$\rho$ statistics}

\begin{figure}[!htb]
    \centering
    \includegraphics[width=1\linewidth]{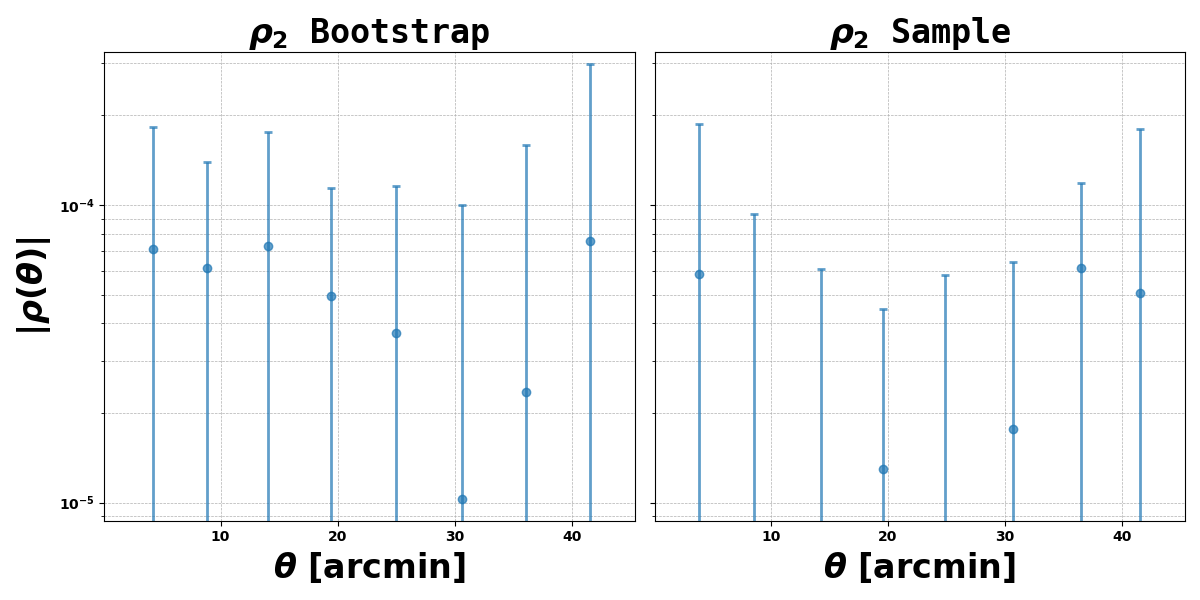}
    \caption{$\rho_2$ correlation function with data taken from the COSMOS-Web point source catalogs. Error bars drawn from bootstrap (left) and sampling (right) approaches. Negative correlations are shown in absolute value, and correlations are plotted in logarithmic scale.}
    \label{fig:rho2comp}
\end{figure}

\begin{figure}[!htb]
    \centering
    \includegraphics[width=1\linewidth]{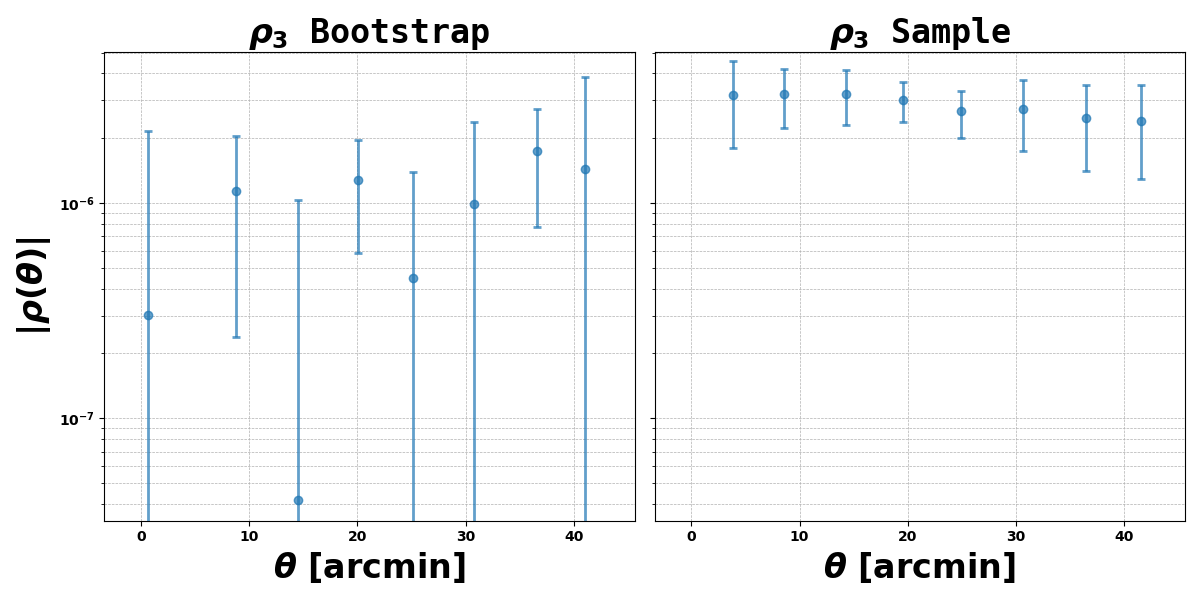}
    \caption{$\rho_3$ correlation function with data taken from the COSMOS-Web point source catalogs. Error bars drawn from bootstrap (left) and sampling (right) approaches. Negative correlations are shown in absolute value, and correlations are plotted in logarithmic scale.}
    \label{fig:rho3comp}
\end{figure}

\begin{figure}[!htb]
    \centering
    \includegraphics[width=1\linewidth]{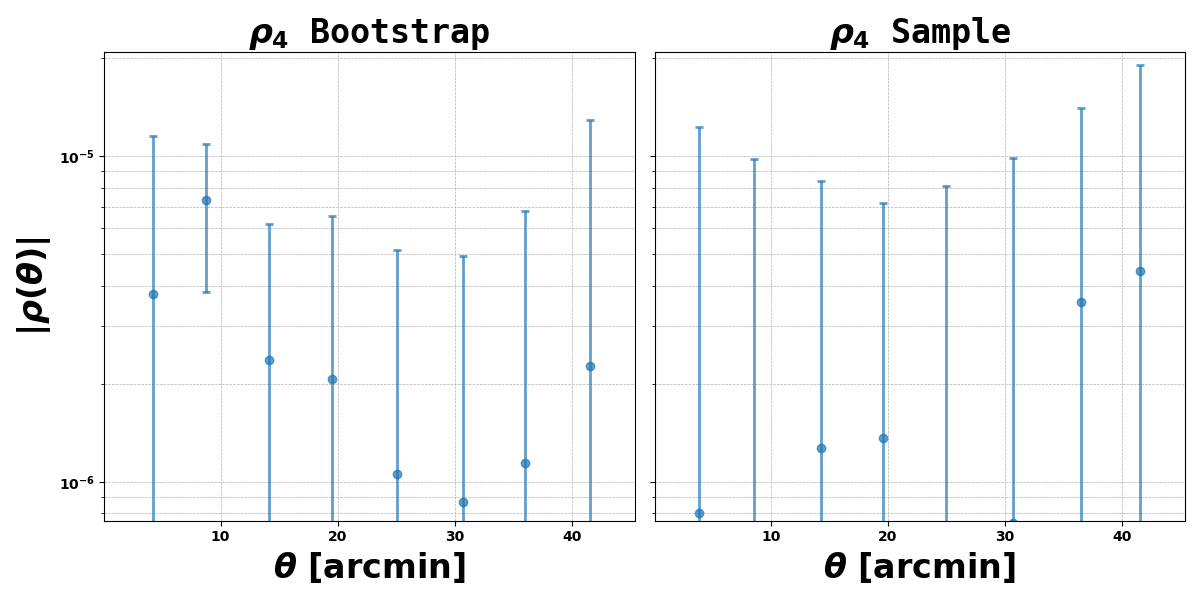}
    \caption{$\rho_4$ correlation function with data taken from the COSMOS-Web point source catalogs. Error bars drawn from bootstrap (left) and sampling (right) approaches. Negative correlations are shown in absolute value, and correlations are plotted in logarithmic scale.}
    \label{fig:rho4comp}
\end{figure}

\begin{figure}[!htb]
    \centering
    \includegraphics[width=1\linewidth]{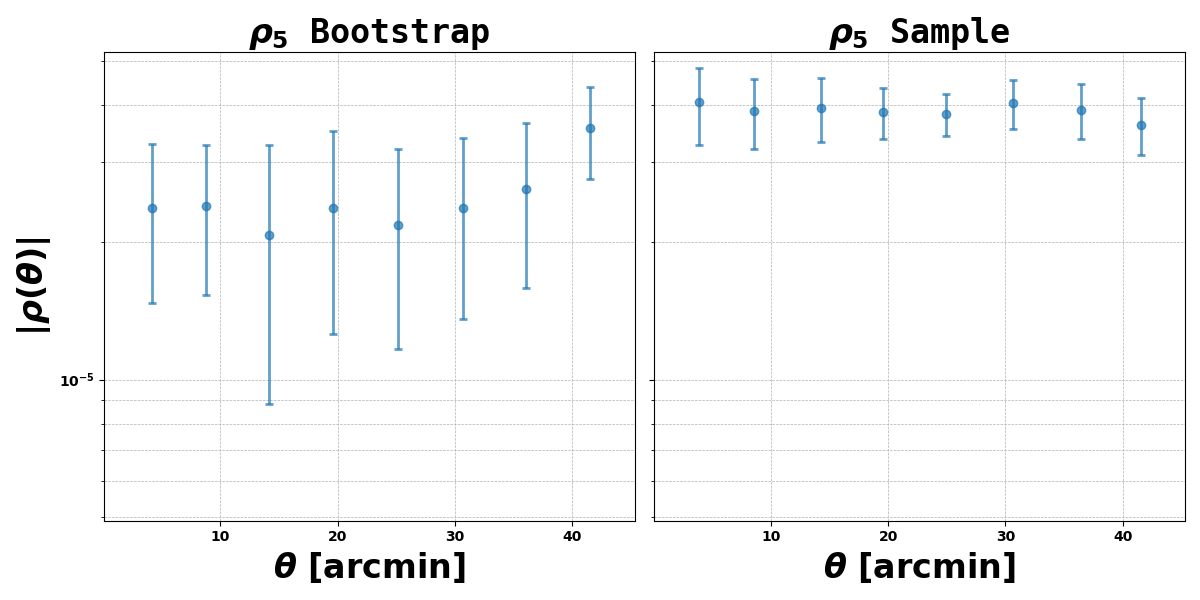}
    \caption{$\rho_5$ correlation function with data taken from the COSMOS-Web point source catalogs. Error bars drawn from bootstrap (left) and sampling (right) approaches. Negative correlations are shown in absolute value, and correlations are plotted in logarithmic scale.}
    \label{fig:rho5comp}
\end{figure}

\clearpage

\section{Jacobians}

\begin{figure}[!htb]
    \centering
\includegraphics[width=0.48\linewidth]{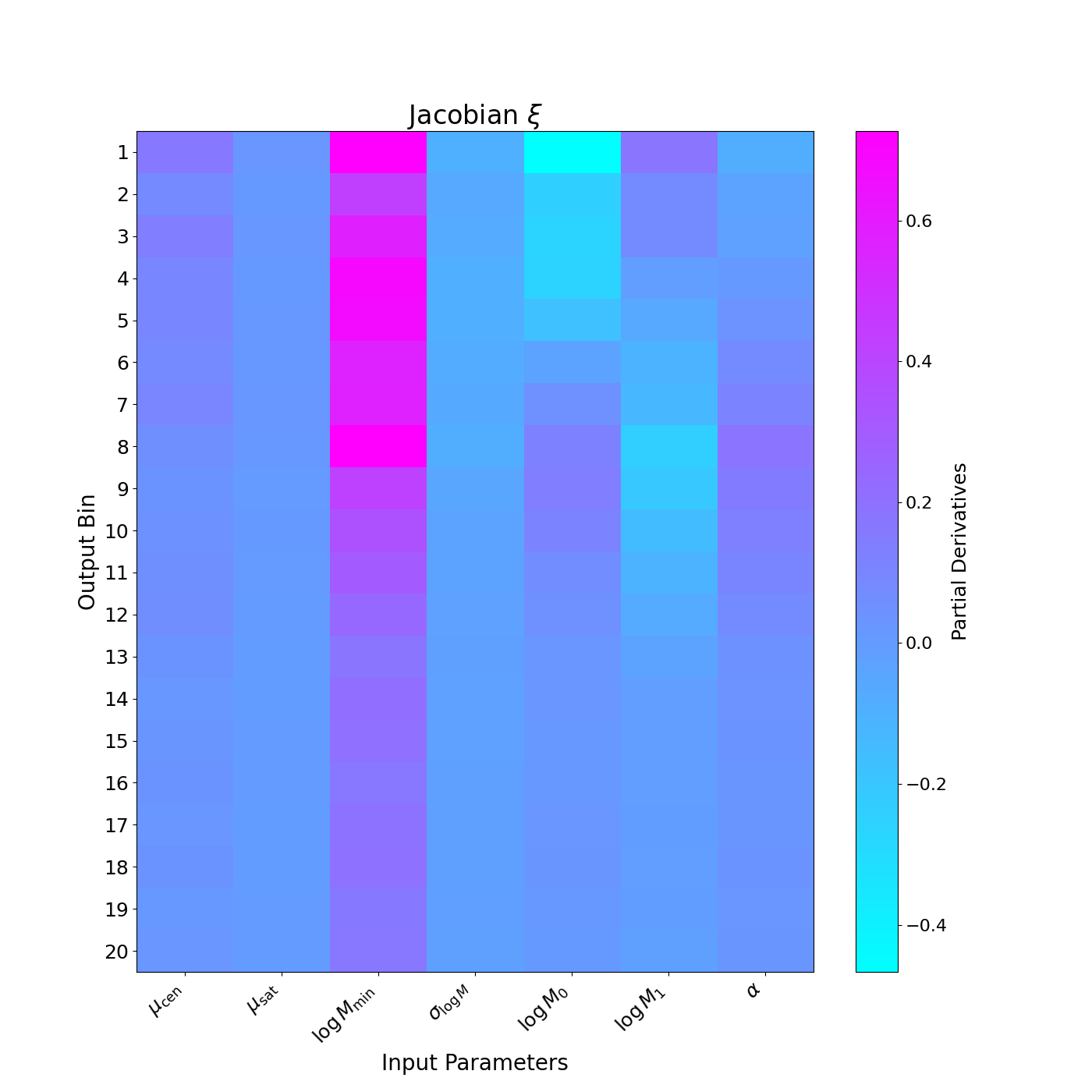}
\includegraphics[width=0.48\linewidth]{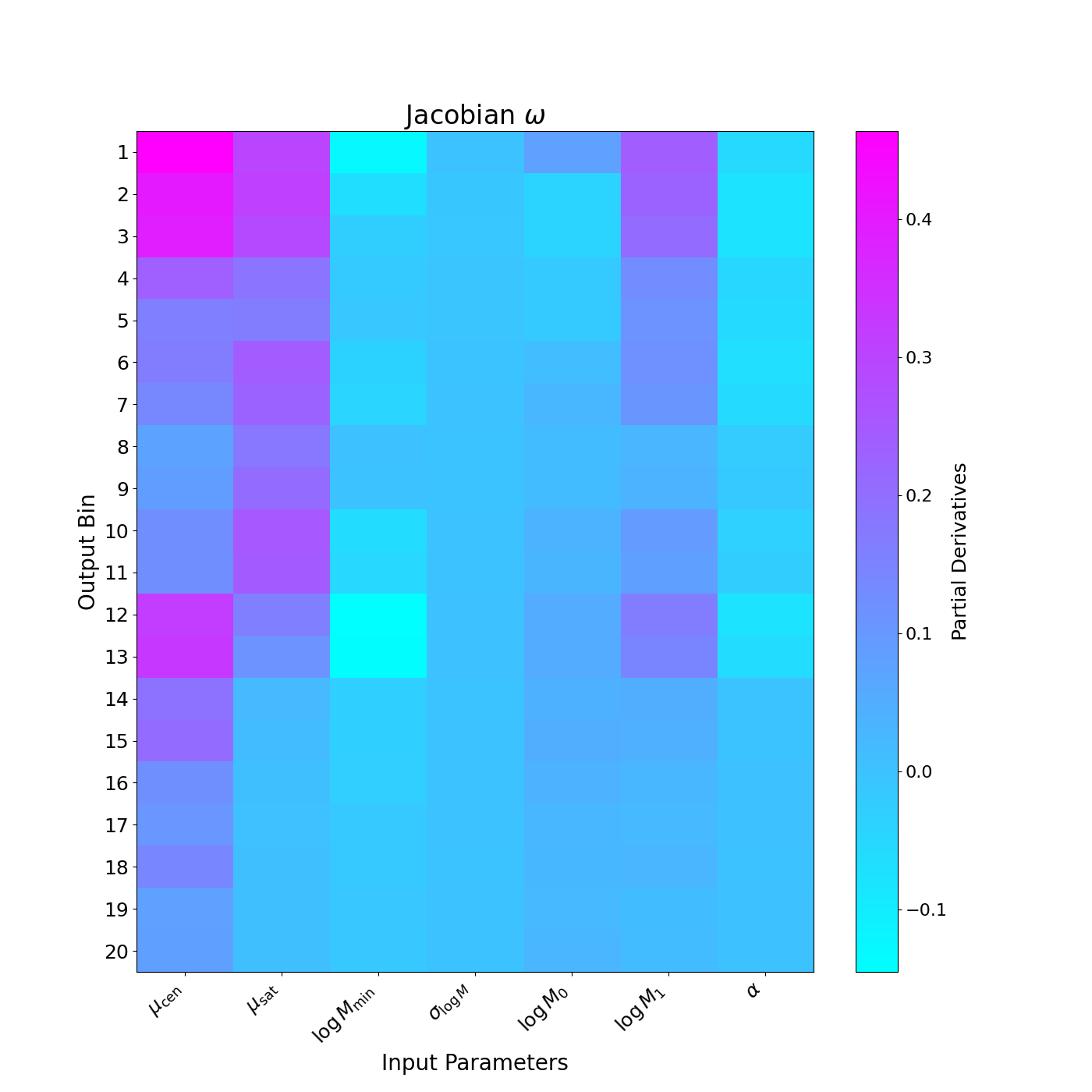}
\includegraphics[width=0.5\linewidth]{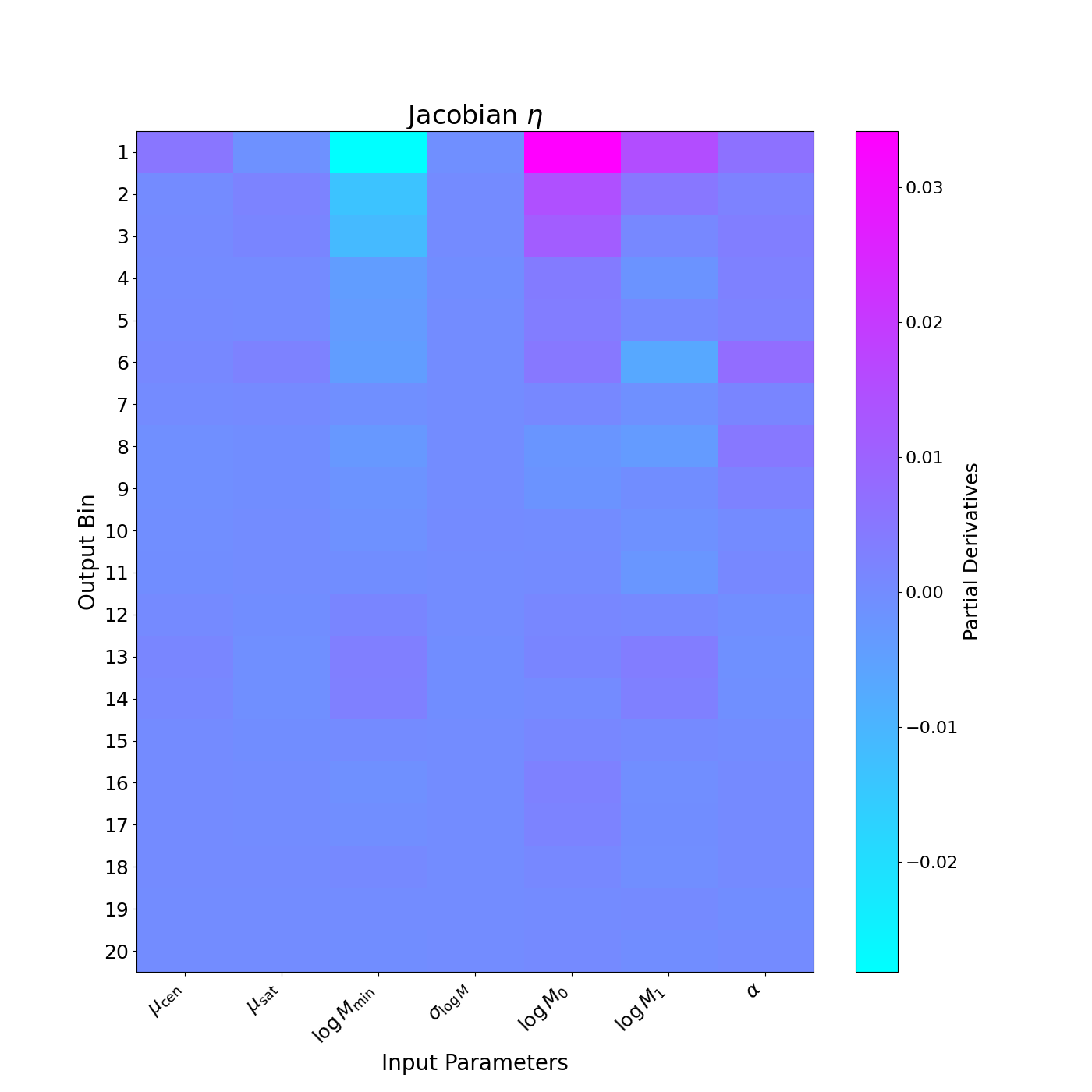}
    \caption{Jacobians associated with each of the $7$ IA parameters and $20$ output bins from IA-Emu.}
    \label{fig:jacobian}
\end{figure}


\end{document}